\documentclass[aps,reprint,twocolumn,superscriptaddress,10pt]{revtex4-2}
\usepackage{siunitx}
\usepackage{amsfonts}
\usepackage{amsmath}
\usepackage{amssymb}
\usepackage{bm}
\usepackage{graphicx}
\usepackage{dcolumn}
\usepackage{mciteplus}
\usepackage{euscript}
\usepackage{multirow}
\usepackage{color}
\usepackage{textcomp}
\usepackage{gensymb}
\usepackage{dblfloatfix}
\graphicspath{{./figs/},{../figs/}}

\DeclareRobustCommand*\cal{\@fontswitch\relax\mathcal}

\def\bfk{{\bf k}}

\def\abinitio{\emph{ab initio}}

\def\nn{\nonumber\\}

\newcommand{\bamnas}{\text{BaMn}_2\text{As}_2}
\newcommand{\srcras}{\text{SrCr}_2\text{As}_2}
\newcommand{\bacras}{\text{BaCr}_2\text{As}_2}
\newcommand{\thcrsi}{\text{ThCr}_2\text{Si}_2}

\newcommand{\ef}{E_{\rm F}}

\newcommand\jij{J_{ij}}

\newcommand{\redbf}[1]{}

\newcommand{\memome}[1]{}

\newcommand{\rfig}[1]{Fig.~\ref{#1}}
\newcommand{\rFig}[1]{Figure~\ref{#1}}
\newcommand{\rtbl}[1]{Table~\ref{#1}}

\begin{document}

\title{Spin dynamics in itinerant antiferromagnet SrCr$_2$As$_2$}
\author{Zhenhua Ning}
\affiliation{Ames National Laboratory, U.S. Department of Energy, Ames, Iowa 50011, USA}

\author{Pinaki Das}
\affiliation{Ames National Laboratory, U.S. Department of Energy, Ames, Iowa 50011, USA}
\affiliation{Department of Physics and Astronomy, Iowa State University, Ames, Iowa 50011, USA}

\author{Y. Lee}
\affiliation{Ames National Laboratory, U.S. Department of Energy, Ames, Iowa 50011, USA}

\author{N. S. Sangeetha}
\affiliation{Ames National Laboratory, U.S. Department of Energy, Ames, Iowa 50011, USA}
\affiliation{Department of Physics and Astronomy, Iowa State University, Ames, Iowa 50011, USA}

\author{Douglas. L. Abernathy}
\affiliation{Quantum Condensed Matter Division, Oak Ridge National Laboratory, Oak Ridge, Tennessee 37831, USA}

\author{D. C. Johnston}
\author{R. J. McQueeney}
\author{D. Vaknin}
\affiliation{Ames National Laboratory, U.S. Department of Energy, Ames, Iowa 50011, USA}
\affiliation{Department of Physics and Astronomy, Iowa State University, Ames, Iowa 50011, USA}
\author{Liqin Ke}
\affiliation{Ames National Laboratory, U.S. Department of Energy, Ames, Iowa 50011, USA}

\date{\today}

\begin{abstract}

SrCr$_2$As$_2$ is an itinerant antiferromagnet in the same structural family as the SrFe$_2$As$_2$ high-temperature superconductors.
We report our calculations of exchange-coupling parameters $J_{ij}$ for SrCr$_2$As$_2$ using a static linear-response method based on first-principles electronic-structure calculations.
We find that the dominant nearest-neighbor exchange coupling $J_{\rm{1}} > 0$ is antiferromagnetic whereas the next-nearest-neighbor interaction $J_{\rm{2}} < 0$ is ferromagnetic with $J_{\rm{2}}$/$J_{\rm{1}}$~=~$-0.68$, reinforcing the checkerboard in-plane magnetic structure. 
Thus, unlike other transition-metal arsenides based on Mn, Fe, or Co, we find no competing magnetic interactions in SrCr$_2$As$_2$, which aligns with experimental findings. 
Moreover, the orbital resolution of exchange interactions shows that $J_1$ and $J_2$ are dominated by direct exchange mediated by the Cr $d$ orbitals.
To validate the calculations we conduct inelastic neutron-scattering measurements on powder samples that show steeply dispersive magnetic excitations arising from the magnetic $\Gamma$ points and persisting up to energies of at least 175 meV. 
The spin-wave spectra are then modeled using the Heisenberg Hamiltonian with the theoretically-calculated exchange couplings.
The calculated neutron-scattering spectra are in good agreement with the experimental data.

\end{abstract}

\maketitle

\section{Introduction}

Magnetic fluctuations have been implicated in the emergence of superconductivity (SC) in iron arsenides as SC often lies in close proximity to a magnetically-ordered state \cite{Johnston2010,Dai2015,Lumsden2010,Inosov2016,Stewart2011,Si2016,Scalapino2012}, similar to observations in high $T_{\rm c}$ cuprates \cite{Vaknin1987,Tranquada1988}. 
In these layered systems, the magnetism is quasi-2D with strong intralayer couplings and weak interlayer couplings. 
Their magnetic ground state is mainly determined by the relative strengths and signs of the nearest-neighbor (NN), and the next-nearest-neighbor (NNN) interactions within the Fe square lattice of the basal plane of the chemical structure \cite{Johnston2010,Dai2015,Lumsden2010}.

Transition metal (TM) arsenides {$AM_2\rm{As}_2$} ($A$ = Ba, Ca, Sr, and $M$ = Cr, Mn, Fe, Co), often crystallize in a body-centered-tetragonal (bct) structure, with a square lattice of the TM ion in the basal plane \cite{Johnston2010,Canfield2010,nedic2023prb}.
Within the $J_1$-$J_2$ model, where $J_1$ denotes the NN and $J_2$ represents the NNN exchange interactions, various magnetic structures have been explored, including collinear commensurate A-type antiferromagnetic (AFM), stripe AFM, G-type AFM, and ferromagnetic (FM) states\cite{Johnston2010}.
For example, stripe-AFM is observed in {$A{\rm Fe}_2\rm{As}_2$} ($A$ = Ca, Sr, Ba) \cite{Johnston2010}, A-type AFM in CaCo$_{2-y}$As$_2$ \cite{Quirinale2013}, and G-type AFM in BaMn$_2$As$_2$ \cite{Singh2009}. 
Inelastic neutron-scattering (INS) studies in these materials have shown that $J_2$ is AFM in nature, which plays an important role in stabilizing unique magnetic structures, such as stripe AFM order. 

The degree of magnetic frustration in $J_1$-$J_2$ system is determined by the ratio $\eta$ = $2J_2 / J_1$, with $\eta = 1$ being maximally frustrated 
\cite{Johnston2010,Dai2015,Shannon2004,Sapkota2017}. 
Some {$AM_2\rm{As}_2$} systems exhibit characteristics of frustrated itinerant antiferromagnetism. 
In some cases, in particular the cobalt arsenides, a high degree of magnetic frustration is present. 
For instance, CaCo$_{2-y}$As$_2$ exhibits an $\eta=0.97$~\cite{Sapkota2017}, while SrCo$_2$As$_2$ has an $\eta=1.78$~\cite{Jayasekara2013}.
This frustration can lead to the emergence of novel states, including spin and electronic nematic \cite{Chandra1990,Shannon2006}, and quantum spin liquids phases~\cite{Balents2010,Savary2016}.
Previous studies on BaMn$_2$As$_2$ have shown that the system is moderately frustrated with $J_2/J_1 \approx 0.33$, where $J_1$ and $J_2$ are both AFM \cite{Johnston2011,Ramazanoglu2017}. 
By doping the BaMn$_2$As$_2$ system with hole carriers, the N\'{e}el temperature remains almost the same up to $x = 0.25$ with minimal effect on the magnetic exchange interactions and the spin-wave spectrum \cite{Ramazanoglu2017}. 
As we show later, a similar trend is found in metallic SrCr$_2$As$_2$ doped with hole carriers.

SrCr$_2$As$_2$ is a member of the magnetic TM arsenides, whose magnetic interactions have not been studied in detail. 
It orders in a G-type AFM structure at a N\'{e}el temperature $T_{\rm N} \approx 591$~K, with a ordered moment of 1.9 $\mu_{\rm B}$/Cr \cite{Das2017}. 
SrCr$_2$As$_2$ closely resembles BaMn$_2$As$_2$ which also displays G-type order with a N\'{e}el temperature $T_{\rm N} \approx 625$~K and a suppressed moment of 3.88 $\mu_{\rm B}$/Mn \cite{Singh2009}. 
Theoretical calculations have shown that in BaMn$_2$As$_2$ and SrCr$_2$As$_2$ there is a strong spin-dependent hybridization between the $3d$ orbitals of Mn or Cr and the As-$4p$ orbitals \cite{An2009,Singh2009b,SinghDJ2009}. 
However, an important difference between BaMn$_2$As$_2$ and SrCr$_2$As$_2$ is that the former is a semiconductor with an intrinsic gap of 0.06~eV \cite{Singh2009b,McNally2015}, while the latter is a metal \cite{Das2017}. 
Thus, a comparison of the properties of the two compounds can shed light on the influences of orbital occupancies and metallic character on the magnetism of SrCr$_2$As$_2$.

Considering the moderate to large frustration found in other {$AM_2\rm{As}_2$} systems, it is of interest to investigate whether SrCr$_2$As$_2$ exhibits characteristics of a frustrated itinerant antiferromagnet since its Cr $3d$ shell is less than half-filled, similar to hole doping in BaMn$_2$As$_2$. 
Later we demonstrate through theoretical calculations that $J_1$ is AFM and $J_2$ is FM in SrCr$_2$As$_2$. 
The absence of magnetic frustration between these interactions firmly stabilizes the checkerboard-like magnetic in-plane structure. 
This is further supported by our INS study where the spin wave (SW) spectrum is reasonably modeled by the calculated magnetic exchange interactions. 
Given the strong link between magnetism and superconductivity, it is important to ascertain whether one can and how to tune magnetic ordering in the system. 
However, our calculation indicates that doping does not induce frustration in SrCr$_2$As$_2$.
Consequently, we conclude that SrCr$_2$As$_2$ is a robust two-dimensional itinerant antiferromagnet, lacking significant magnetic frustration.

\section{Methods}\label{sec:methods}

\subsection{Crystal structure}

\begin{figure}[h]
\centering
\begin{tabular}{c}
\includegraphics[width=.65\linewidth,clip]{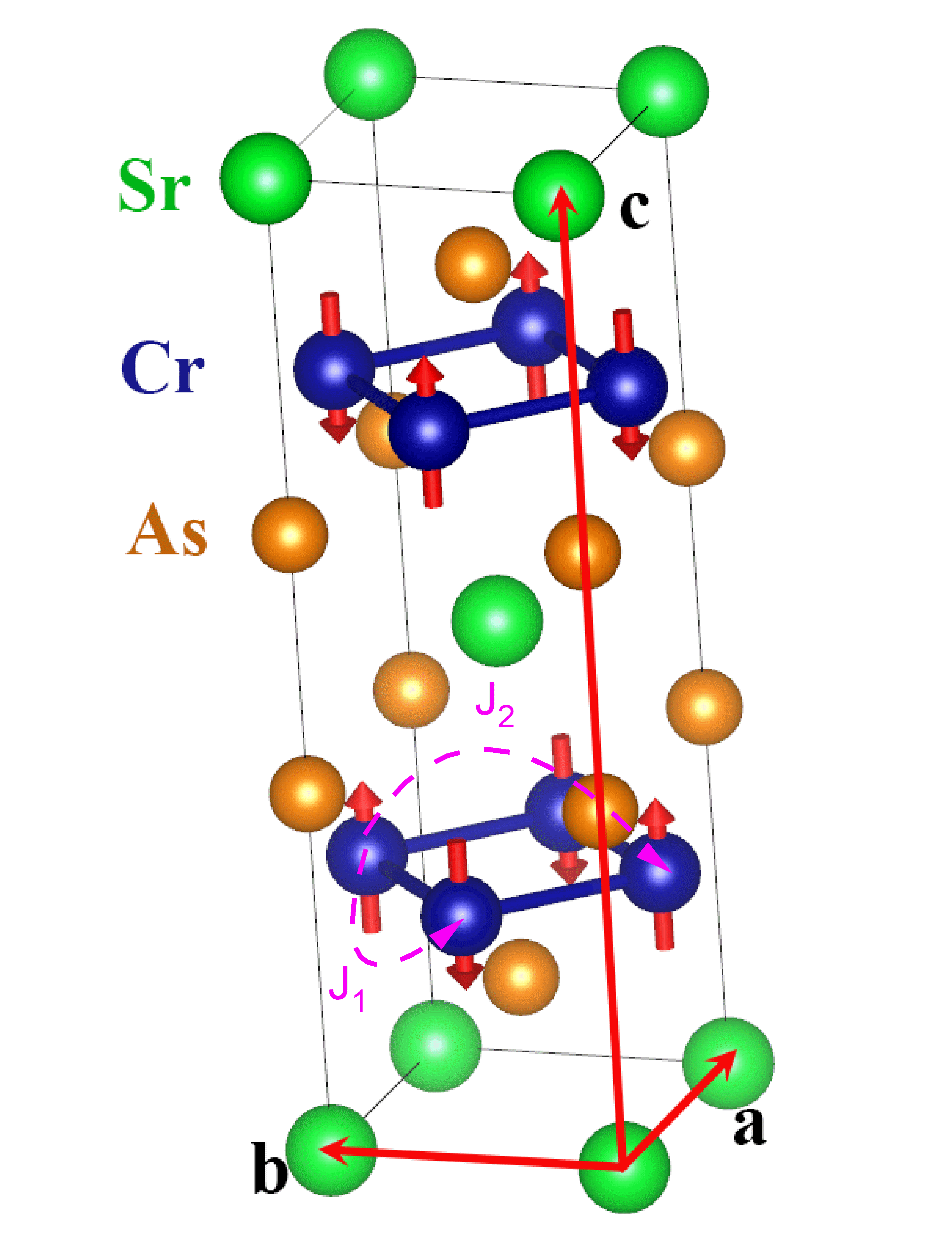} \\
\end{tabular}%
\caption{
Schematic representation of the crystal structure of $\srcras$.
Sr, Cr, and As atoms are represented by green, blue, and orange spheres, respectively.
The lattice vectors $\bold{a}$, $\bold{b}$, and $\bold{c}$ are highlighted in red.
The longest lattice vector $\bf c$ along the $\hat{z}$ direction, and Cr atoms are arranged in a square net with N\'{e}el AFM order on the ${\bold a}{\bold b}$-basal plane, or equivalently, the ${xy}$-plane.
The G-type magnetic structure is also shown.
$J_1$ and $J_2$ denote the NN exchange coupling and NNN exchange coupling, respectively. 
}
\label{srcras_structure}
\end{figure}

	$\srcras$ crystallizes in the tetragonal $\thcrsi$-type structure (I4/mmm, space group no. 139), as depicted in in \rfig{srcras_structure}. The primitive cell of this structure consists of one formula unit (f.u.), while the conventional cell contains two f.u.
	The Cr atoms occupy the $4d$(-4m2) sites and form a square net in the $\bf ab$ plane with G-type AFM ground state, as shown in \rfig{srcras_structure}.
	The As atoms occupy the $4e$(4mm) sites, forming edge-sharing Cr$_4$As tetrahedra with neighboring Cr atoms. 
	Each Sr atom, which occupies the $2a$(4/mmm) site, is surrounded by eight As atoms.
	Together they form a body-centered-tetragonal unit cell elongated along the $\bold c$ axis.
	Lying in the $\bf ab$ plane, the nearest Cr-Cr bond has a length of $\SI{2.76}{\angstrom}$.
	In the Sr-As cage, the length of the Sr-As bond is of  $\SI{3.26}{\angstrom}$ and Sr-Sr bond is of $\SI{3.91}{\angstrom}$ in length.	
	All calculations in this work use experimental lattice parameters $a=b=\SI{3.91}{\angstrom}$, $c=\SI{12.93}{\angstrom}$ and atomic positions $z_{\rm{As}}$ = 0.3667 obtained at $T=12$K~\cite{Das2017}.

\subsection{Computational Details}

We conduct density functional theory (DFT) calculations using the full-potential linear augmented plane wave (FP-LAPW) method as implemented in the \textsc{wien2k} package~\cite{Blaha2020,wien2k} with both the local density approximation (LDA), with the Barth-Hedin parametrization~\cite{Barth1972}, and the generalized-gradient approximation (GGA)~\cite{Perdew1996} exchange-correlation potentials. 
To facilitate our subsequent analyses, we construct the maximally localized Wannier functions (MLWFs) through a postprocessing procedure~\cite{marzari1997prb, souza2001prb, marzari2012rmp}, as implemented in \textsc{wannier90}~\cite{mostofi2014cpc}, using the output of the self-consistent scalar-relativistic DFT calculation.
In total, the tight-binding (TB) basis comprises 45 MLWFs, encompassing $s$-, $p$-, and $d$-orbitals for all five atoms within the unit cell.
To minimize the spread functional for entangled energy bands, we adopt a two-step procedure~\cite{souza2001prb}.
For each spin channel, a real-space Hamiltonian $H^\sigma(\bf{R})$ with dimensions 45$\times$45 is constructed to accurately represent the band structures in a specified ``frozen'' energy window near the Fermi energy $E_{\text{F}}$.
The energy bands are recalculated within TB to ensure that DFT bands can be accurately reproduced before further magnetic-property calculations.

The pairwise exchange-coupling parameters $J_{ij}$ are calculated using a static linear-response approach~\cite{Liechtenstein1984}, which was implemented in our recently-developed TB code, that has been employed to efficiently analyze band structures~\cite{rosenberg2022prb,lee2023prb}, Fermi surfaces~\cite{timmons2020prr}, magnetocrystalline anisotropy~\cite{ke2019prb}, and ferrimagnetic alignment~\cite{lukashev2023jap}.
Further details regarding the method and its applications to other materials can be found elsewhere~\cite{Ke2013,Ke2017}.
To calculate exchange parameters $J({\bf q})$, we construct the intersite Green's function $G_{ij}^{\sigma}(\bfk)$ on a $32\times 32\times 32$ $k$ mesh in the Brilliouin zone. 
The real-space exchange constants $J ({\bf R})$ are obtained through a subsequent Fourier transform of $J({\bf q})$. 
The theoretical methods introduced above align with those presented in Refs.~\cite{Ning2024prb}, and additional details are provided herein.
All $J_{ij}$ discussed hereafter, unless specified otherwise, are the same as the $J_{ij}^\text{N}$ in the above reference.

The magnetic neutron-scattering intensity, $S$({\bf Q},$E$), can be calculated by solving the linear spin-wave equations in the Heisenberg model~\cite{Johnston2011,Ramazanoglu2017}. 
The Hamiltonian for the $J_1$-$J_2$-$J_c$ Heisenberg model is given by
\begin{equation}
\begin{split}
H ~=~ & J_1\sum_{{\rm NN},\bf a \bf b} {\rm \bf S}_i \cdot {\rm \bf S}_j ~+~ J_2\sum_{{\rm NNN},\bf a \bf b} {\rm \bf S}_i \cdot {\rm \bf S}_j \\
 & ~+~ J_c\sum_{{\rm NN},\bf c} {\rm \bf S}_i \cdot {\rm \bf S}_j~,
\label{eq:1}
\end{split}
\end{equation}
where $J_1$ and $J_2$ are the NN and NNN magnetic exchange parameters in the square $\bf a \bf b$ basal plane of the tetragonal lattice and $J_c$ is the NN interlayer exchange coupling along the $\bf c$-direction. 
Throughout this paper, the magnetic exchange parameter $J$ is considered positive for AFM interactions and negative for FM couplings.
The degree of magnetic frustration in a $J_1$-$J_2$ system is characterized by the ratio, $\eta$ = $2J_2 / J_1$, with $\eta = 1$ being maximally frustrated 
\cite{Johnston2010,Dai2015,Shannon2004,Sapkota2017}. 
For materials with G-type AFM magnetic ordering, an antiferromagnetic $J_2$ leads to a frustration and a stripe-type magnetic ordering when $2J_2 / J_1>1$, while a ferromagnetic $J_2$ reinforces the magnetic ordering. 
The ground state of a system is a G-type magnetic ordering when both $J_1$ and $J_c$ are antiferromagnetic~\cite{Johnston2011,Ramazanoglu2017}. 
Note that the sign of $J_c$ distinguishes between G-type and C-type magnetic structures where both have N\'{e}el type in-plane ordering.

With the obtained $\abinitio$ exchange parameters, the spin-wave (SW) spectra are calculated through the bosonization of the Heisenberg Hamiltonian with Holstein-Primakoff transformation~\cite{holstein1940pr,Johnston2011,Ramazanoglu2017}, as given by  
\begin{equation}
\begin{split}
\frac{1}{4} \left [ \hbar \omega ({\rm \bf q}) \right ]^2 = & \left [ 2SJ_1 + SJ_c - SJ_2 (2 - {\rm cos}~q_xa -{\rm cos}~q_ya) \right ]^2 \\
& -\bigg \{ SJ_1 \left [ {\rm cos}\frac{(q_x + q_y)a}{2} + {\rm cos}\frac{(q_x - q_y)a}{2} \right ] \\
& + SJ_c~{\rm cos}\frac{q_z c}{2}\bigg \}^2~,
\label{eq:3}
\end{split}
\end{equation}
where $a$ and $c$ are the lattice parameters for the bct unit cell and $\mathbf{q}=(q_x,q_y,q_z)$ is the wave vector measured relative to the G-type magnetic Bragg peak. 
In the absence of an anisotropy-induced energy gap, the spin-wave dispersion in a bct unit cell can be expressed as
\begin{equation}
\hbar \omega ({\rm \bf q}) = \hbar \sqrt{v_{ab}^2 (q_x^2 + q_y^2) + v_c^2 q_z^2}~,
\label{eq:4}
\end{equation}
where the spin-wave velocities $v_{ab}$ in the $ab$-plane and $v_c$ along the $c$-axis are given by
\begin{eqnarray}
\hbar v_{ab} &=& 2J_1 S a \sqrt{\left( 1 - \dfrac{2 J_2}{J_1}\right)\left( 1 + \dfrac{J_c}{2J_1}\right) }\,,  \nn
\hbar v_{c} &=& \sqrt{2} J_1 S c \sqrt{\dfrac{J_c}{J_1}\left( 1 + \dfrac{J_c}{2J_1}\right) }\,.
\end{eqnarray}

\subsection{Experimental Details}

A polycrystalline sample of SrCr$_2$As$_2$ was synthesized by solid-state reaction and characterized in accordance with the methods described in Ref.~[\citenum{Das2017}]. 
INS measurements were carried out on the ARCS spectrometer at the Spallation Neutron Source located at Oak Ridge National Laboratory. 
The polycrystalline sample with a mass of 2 grams was placed in a cylindrical aluminum (Al) sample can and mounted on the cold tip of a closed-cycle He cryostat. 
The measurements were performed at $T = 10$ K using incident energies $E_i = 50, 150, 300, 500$, and $1000$ meV.
The energy-dependent resolution is approximately $3-5\%$ of the full width at half maximum (FWHM) of $E_i$. 
The data were corrected for both Al (sample holder) and hydrogen (H) scattering (due to water adsorption on the polycrystalline sample's surface resulting from exposure to air). 
Incoherent nuclear scattering from a vanadium standard was used to normalize the detector efficiency across the detector bank. 
The dynamical structure factor $S({\textbf{Q}},E)$, where \textbf{Q} represents the scattering vector, and the corresponding momentum cuts were then obtained using MSLICE software~\cite{Azuah2009}.

\section{Results and Discussion}

\subsection{Exchange coupling from DFT calculations}~\label{sec:theoresults}

\begin{table}[bhtp]
  \caption{ The values of on-site Cr atomic magnetic moment $\mu_\text{Cr}$~($\mu_B$/atom) and the magnetic energy $E_\text{mag}$ (meV/Cr) for various spin configurations in $\srcras$ calculated using the DFT method.
The spin configuration denoted as Stripe-AFM (FM) represents a stripe-type ordering within the $\bf ab$-plane with AFM (FM) coupling along $\bf c$-axis.
The magnetic energy is calculated with respect to the energy of the nonmagnetic structure.}
\label{tbl:energy}%
\begin{tabular*}{\linewidth}{l @{\extracolsep{\fill}} ccrr}
  \hline\hline
  \\[-1em]
   \multicolumn{1}{c}{Spin}            & \multicolumn{2}{c}{$\mu_\text{Cr}$} & \multicolumn{2}{c}{$E_\text{mag}$}  \\ \cline{2-3} \cline{4-5}
   Configuration   &   LDA   & GGA    &   LDA       &   GGA    \\
  \\[-1.1em]
  \hline
  \\[-0.5em]
  $G$-type         &  1.96   &  2.23  &   -171.13   &  -277.67 \\
  $C$-type         &  2.02   &  2.30  &   -165.35   &  -277.35 \\
  $A$-type         &  1.17   &  1.69  &    -40.60   &   -93.43 \\
  FM               &  1.22   &  1.57  &    -47.76   &   -92.09 \\
  Stripe-AFM       &  1.59   &  2.06  &     -1.54   &   -82.48 \\
  Stripe-FM        &  1.54   &  2.03  &     -2.11   &   -79.01 \\
  \\[-0.7em]
  \hline\hline
\end{tabular*}
\end{table}

\rtbl{tbl:energy} presents the DFT results for the on-site Cr moment and magnetic energy in various spin configurations, including FM, N\'eel, and stripe in-plane magnetic orderings. 
The calculations confirm that $G$-type AFM ordering is the most stable ordering in $\srcras$, favoring AFM coupling along the $z$ direction. 
Interestingly, $C$-type and $G$-type spin configurations compete for the ground state. 
Their energy difference is small and depends on the exchange functional used in the calculations. 
In the Hamiltonian in Eq.~\eqref{eq:1}, this indicates that $J_c$ is relatively small and the system is quasi-2D.
The experimental ordered moment for $\srcras$ is determined to be $\mu_\text{Cr}=1.9~\mu_\text{B}$ at $T$ = \SI{12}{\K}~\cite{Das2017}, which is similar to the measured value in $\bacras$~\cite{Filsinger2017}. 
Within the LDA, the DFT yields $\mu_\text{Cr}=1.96~\mu_\text{B}$, which agrees well with experimental findings.
Conversely, the GGA gives $\mu_\text{Cr}=2.23~\mu_\text{B}$, slightly overestimating the on-site Cr moment.
This indicates that the spin magnetic moments observed in both experiments and calculations are significantly smaller than the $4.0~\mu_\text{B}/\text{Cr}^{2+}$ predicted in an insulator.
Notably, DFT calculations suggest that the Fermi level crosses the Cr-d bands, partially occupying the 3d orbitals in both spin channels and a non-integer spin magnetic moment. 
Overall, our calculations imply itinerant magnetism in $\srcras$, consistent with previous neutron diffraction studies~\cite{Das2017}.

For simplicity, one can determine the strong in-plane exchange-coupling parameters by mapping the DFT total energy to the model Heisenberg Hamiltonian.
In the case of metallic systems including $\srcras$, however, the approach becomes challenging. 
It is known that the Heisenberg Hamiltonian assumes fixed magnetic moments.
As demonstrated in Table \ref{tbl:energy}, the on-site Cr moment exhibits different values among different in-plane spin configurations, varying significantly from $1.17~\mu_\text{B}$ to $2.02~\mu_\text{B}$. 
Therefore, extracting reliable exchange parameters in $\srcras$ using the total-energy approach becomes questionable.
To address this issue, we employ a linear response approach to calculate the exchange parameters $J_{ij}$. 
This approach offers a more suitable way for determining exchange interactions in metallic systems.
Additionally, considering that the LDA results align better with experimental observations than the GGA, we will focus solely on the calculations performed within the LDA framework.

\begin{figure}[htbp]
\includegraphics[width=1.00\linewidth,clip]{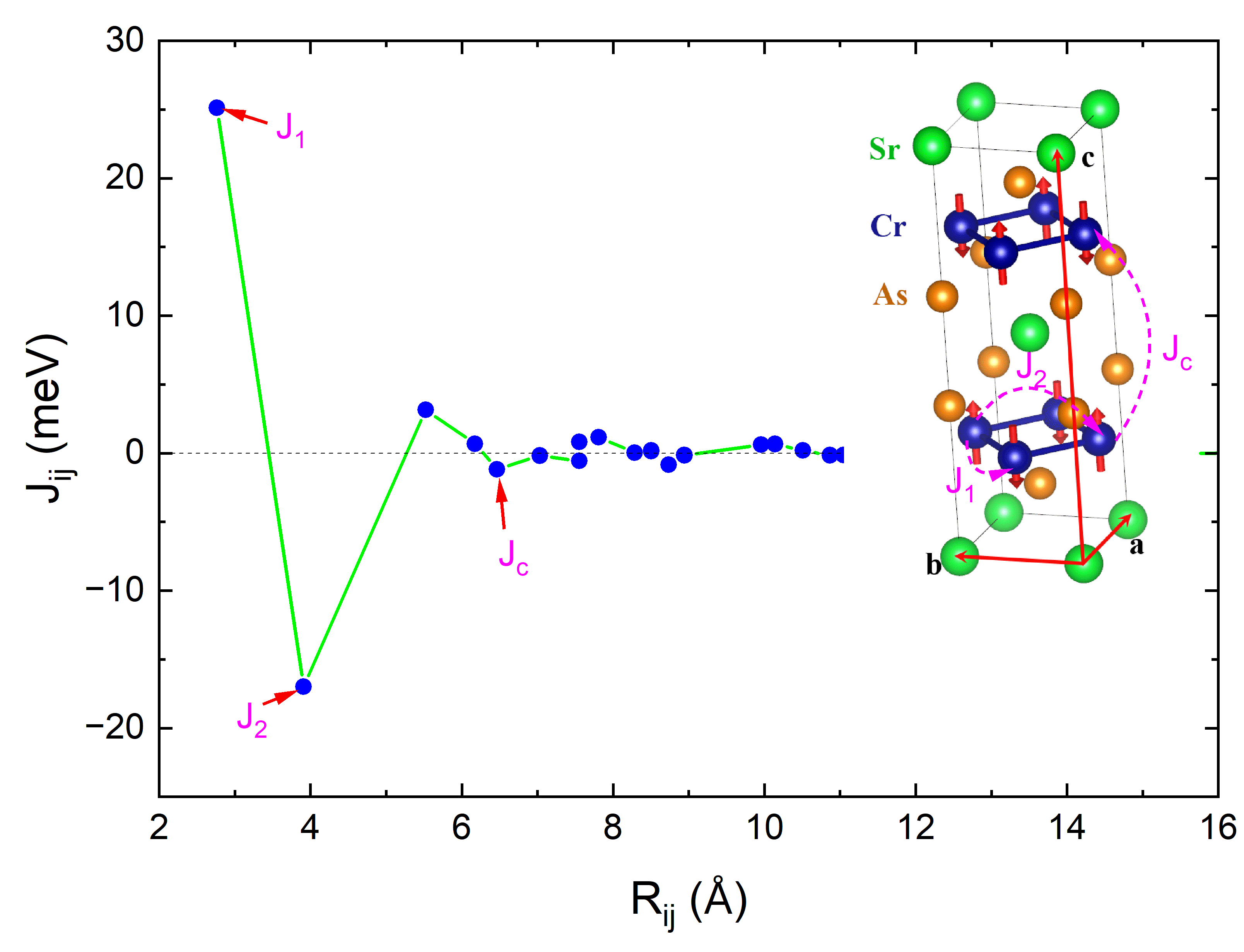}
\caption{(Color online) Real-space pairwise magnetic exchange parameters $\jij$ as a function of distance (in the unit of \AA) between two Cr atoms in G-type $\srcras$. 
$\jij$ is defined in Eq.~\eqref{eq:1}, and calculated using a linear response method.   
Clearly, $J_{1}$ is AFM and $J_{2}$ is FM in $\srcras$.}
\label{Fig:jij}
\end{figure}

In Figure \ref{Fig:jij}, the calculated exchange parameters are illustrated as a function of the distance $R_{ij}$ between the Cr atoms.
They are dominated by the NN exchange coupling $J_1$ and NNN exchange coupling $J_2$, showing a rapid decay as $R_{ij}$ increases and becoming negligible after $R_{ij}=\SI{6}{\AA}$.
Notably, the NN exchange $J_1$ is positive, indicating an AFM interaction, while the NNN $J_2$ is negative, suggesting a FM interaction.
This observation indicates the absence of spin frustration in $\srcras$.
Furthermore, the calculated ratio $J_{2}/J_{1}=-0.68$ is comparable to the previously reported value of $J_2$/$J_1=-0.85$ to $-0.51$ for $\bacras$~\cite{SinghDJ2009,Zeng2017prb,Zhou2019jmmm,Hamri2021jpcs}.
It is worth noting that the previously reported ratios for $\bacras$, obtained through fitting the total energy of various spin configurations to the Heisenberg Hamiltonian, vary widely due to different magnetic moments used in the calculations.
This wide range of values further supports the above mentioned challenges associated with the application total-energy mapping approach in metallic systems.

\begin{figure}[ht]
\centering
\begin{tabular}{c}
\includegraphics[width=0.9\linewidth,clip]{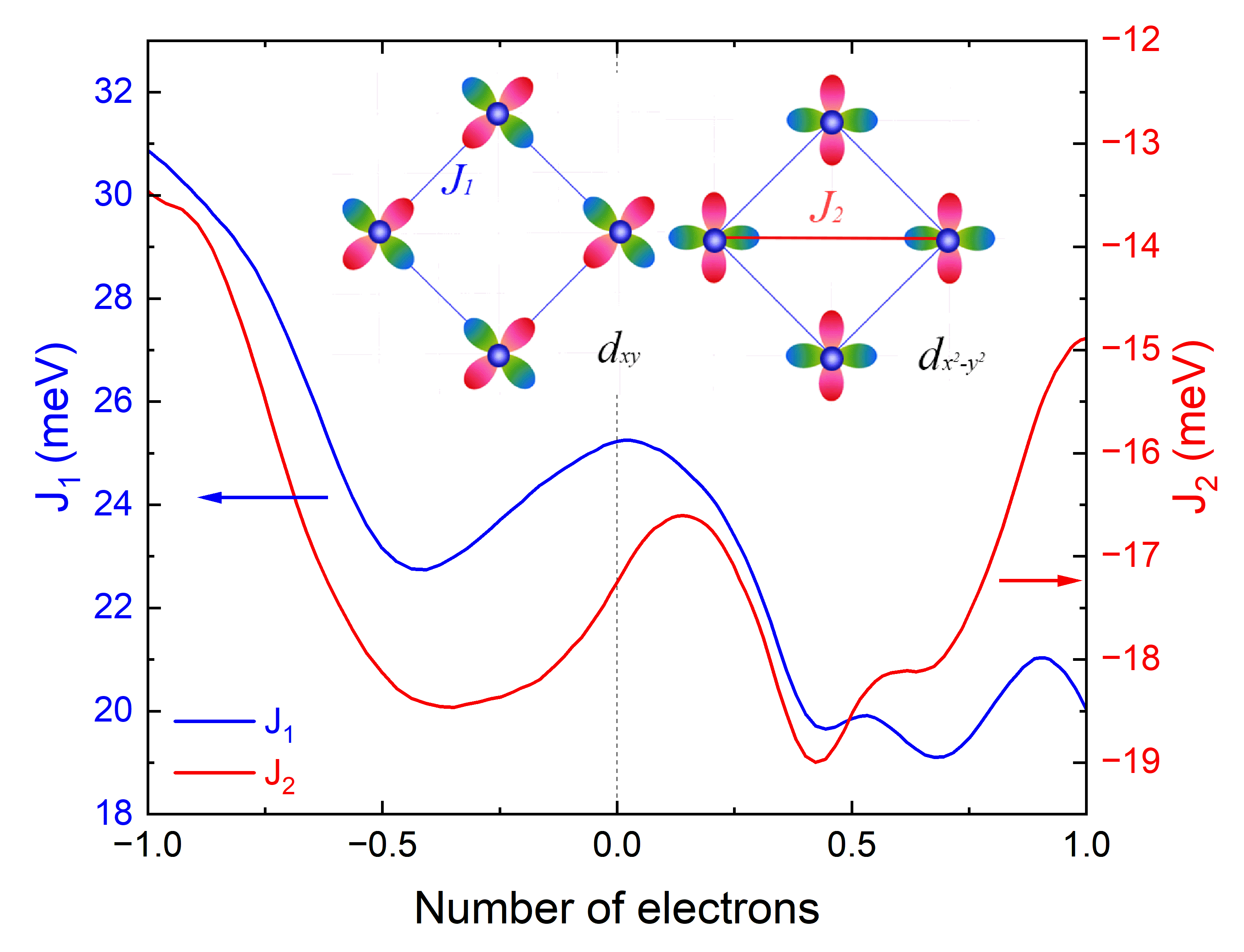}   \\
\end{tabular}%
\caption{ Exchange parameters $J_1$ and $J_2$ of one Cr site of $\srcras$ 
  as functions of bandfilling. 
  The Fermi level is shifted to zero. 
  It shows $J_1$ and $J_2$ as functions of bandfilling or electron and hole doping in units of the number of electron (hole) per f.u.
The insets illustrate top views of $d_{xy}$ orbitals and $d_{x^2-y^2}$ orbitals that connect Cr (in blue spheres) atoms.
$J_1$ and $J_2$ denote the NN and NNN exchange couplings.
}
\label{fig:j12cd}
\end{figure}

\begin{table}[htbp]
  \caption{Pairwise exchange parameters $\jij$ for the NN and NNN exchange parameters $J_1$ and $J_{2}$ (\si{meV}) and their contributions from Cr-$3d$ orbitals in $\srcras$.
  The second column presents the degeneracy of the exchange parameter.
}
\label{tbl:j12_orb}
\bgroup
\def\arraystretch{1.1}
\begin{tabular*}{\linewidth}{c@{\extracolsep{\fill}}cccccccr}
\hline\hline  
$\jij$   & No. & $R_{ij}$  & $xy$ & $yz$ & $z^2$ &  $xz$ & $x^2-y^2$ & Total  \\ \hline
$J_1$    & 4   &  2.77  & 13.75 & 1.56 & 0.04 & 1.56 & 6.12 & 25.1 \\
$J_{2}$ & 4   & 3.92 &  -0.88  & -2.07 & -5.94 & -2.07  & -6.43 &  -17.0 \\
\hline
\end{tabular*}
\egroup
\end{table}
The values of $J_1$ and $J_{2}$, along with their Cr-$3d$ orbital contributions, are presented in \rtbl{tbl:j12_orb}.
For the definition and calculation of the orbital contributions to the exchange couplings, refer to the Ref.~\cite{Ning2024prb}.
The dominant contribution to the AFM coupling $J_1$ originates from the $d_{xy}$ orbital, which directly connects two NN Cr sites, as depicted in the inset of Fig.~\ref{fig:j12cd}.
Moreover, the $d_{x^2-y^2}$ orbital significantly contributes to $J_1$, while other $d$ orbitals make smaller AFM contributions to $J_1$. 
The primary contribution to FM $J_2$ arises from the $d_{x^2-y^2}$ orbital, that points along the $\mathbf{a}$/$\mathbf{b}$ direction, as depicted in the inset of Fig.~\ref{fig:j12cd}. 
The $d_{z^2}$ orbital contributes a similar magnitude to $J_2$, indicating a significant role of the indirect superexchange through As layers adjacent to the Cr layer. 
Additionally, the $d_{yz}$ and $d_{xz}$ orbitals exhibit the same contributions to both $J_1$ and $J_{2}$, reflecting the symmetry in the square Cr basal plane.
The effective interlayer exchange coupling $J_{c} = 1.2$ meV is notably weaker owing to the substantial distance between Cr layers along the $\mathbf{c}$ ($\hat{z}$) direction. 
Our calculations show that carrier doping does not induce a significant change in $J_{c}$, making it constantly weak compared to $J_1$ and $J_2$. 

\rFig{fig:j12cd} illustrates the dependence of the exchange parameters $J_1$ and $J_2$ on doping carriers.
With hole doping up to 1 hole per unit cell, the amplitudes of $J_1$ and $J_2$ change simultaneously and roughly balance each other, which leads to no significant change in the N\'{e}el temperature.
This suggests a similar behavior in $\srcras$ to what has been experimentally observed in $\bamnas$, where the N\'{e}el temperature remains nearly constant up to $x = 0.25$~\cite{Ramazanoglu2017}.
In the case of electron doping, up to 0.2 electrons per unit cell, the amplitudes of both $J_1$ and $J_2$ decrease, resulting in a reduction in the N\'{e}el temperature.
For electron doping between 0.2 and 0.4 electrons per unit cell, the amplitudes of $J_1$ and $J_2$ change simultaneously and roughly balance each other, leading to  no significant change in the N\'{e}el temperature within this range.
However, beyond electron doping of 0.4 electrons per unit cell, the amplitude of $J_2$ increases significantly while the amplitude of $J_1$ changes only slightly, causing a decrease in the N\'{e}el temperature.
Overall, electron doping tends to decrease the N\'{e}el temperature.
We note that $J_1$ and $J_2$ exhibit similar trends as the change of doping carriers, a similar behavior was also observed in SrMnSb$_2$~\cite{Ning2024prb}. 
Meanwhile, we predict that magnetic frustration does not emerge in $\srcras$ because $J_2$ remains negative with carrier doping, as depicted in Fig.~\ref{fig:j12cd}.

\begin{figure}[!ht]
\centering
\begin{tabular}{c}
\includegraphics[width=0.93\linewidth,clip]{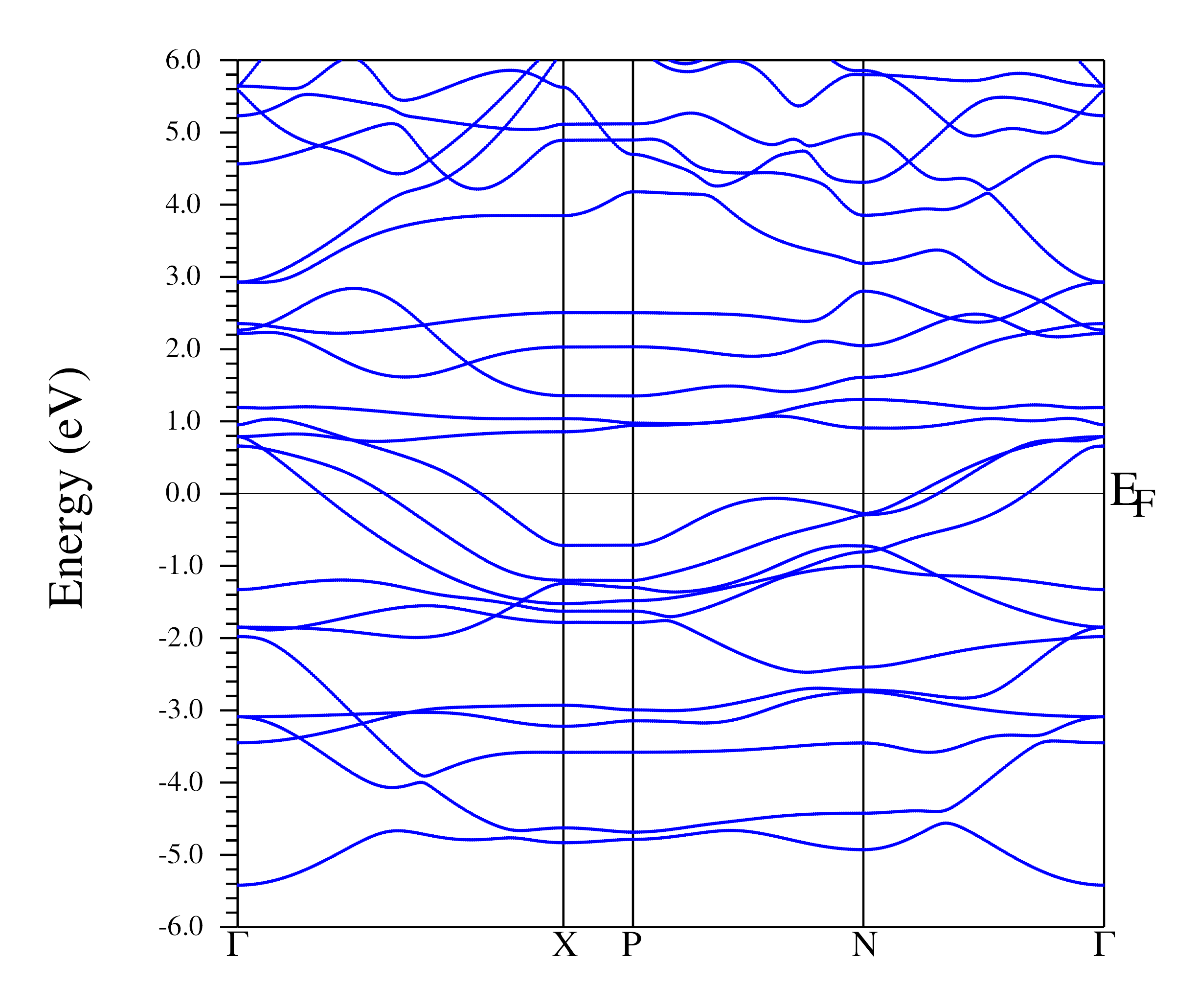} \\  
\includegraphics[width=.8\linewidth,clip]{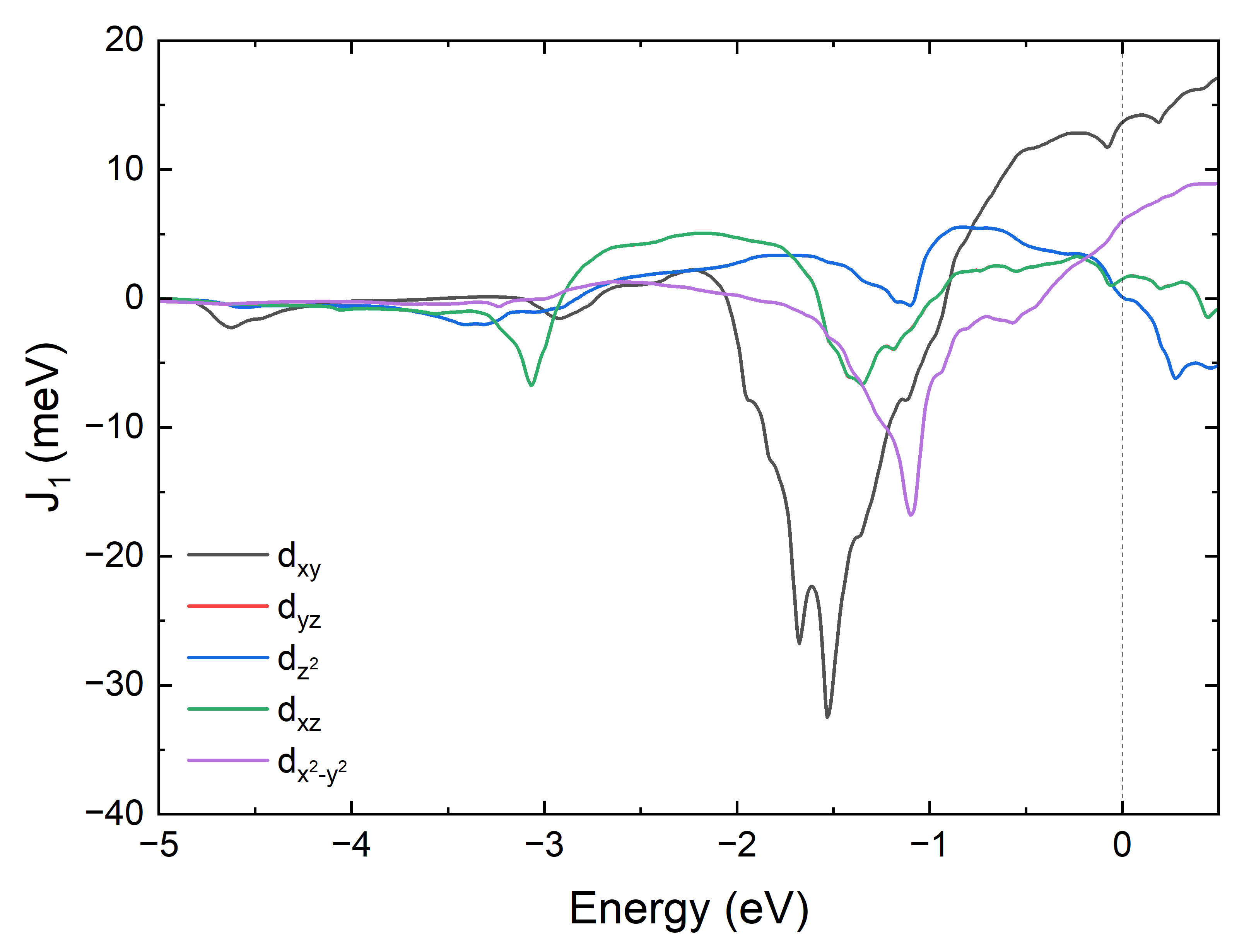} \\  
\includegraphics[width=.8\linewidth,clip]{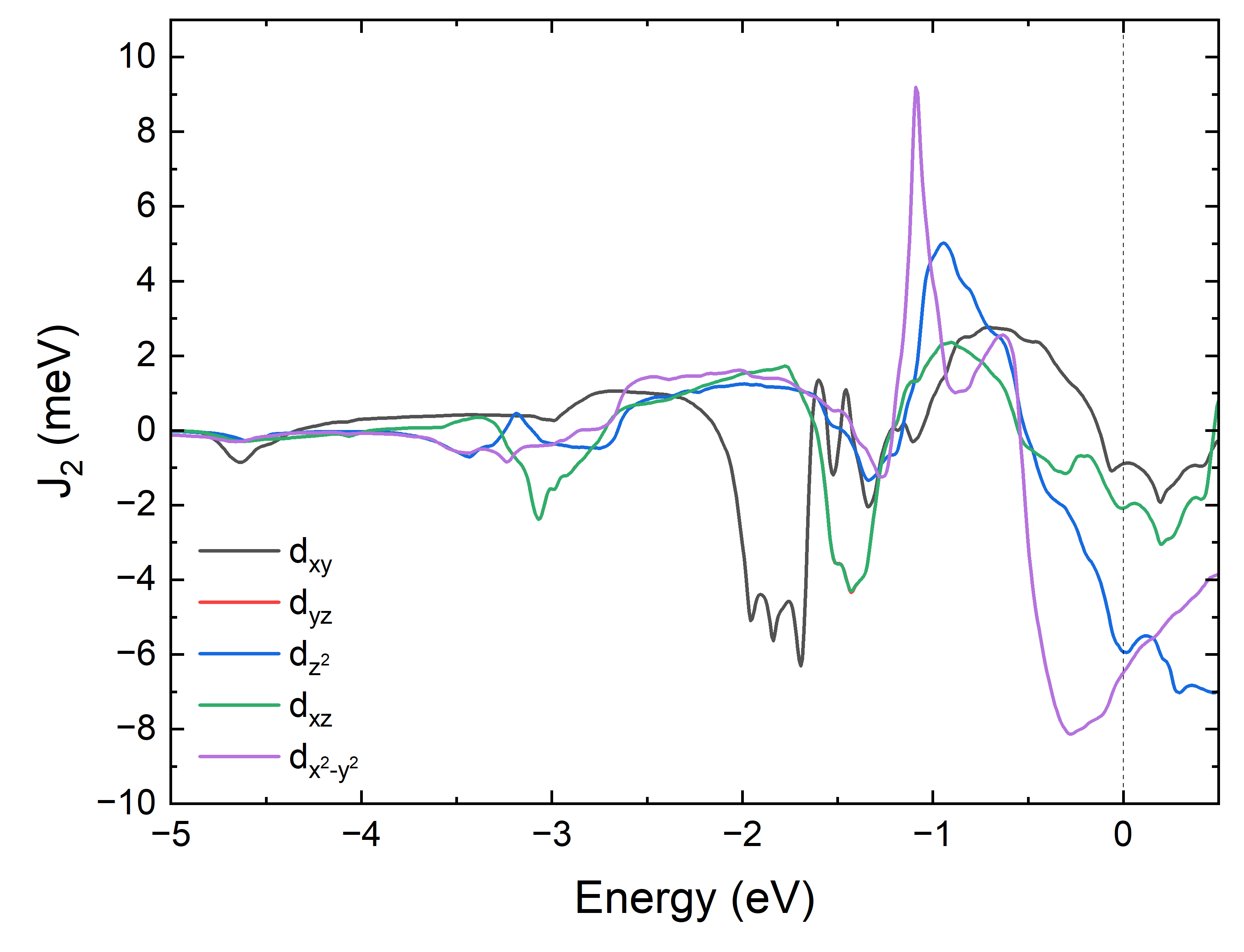} \\  
\end{tabular}%
\caption{ Band structure for SrCr$_2$As$_2$ and exchange parameters $J_1$ and $J_2$ of one Cr site of $\srcras$ 
  as functions of bandfilling. 
  The Fermi level is shifted to zero. 
  It shows $J_1$ and $J_2$ as functions of bandfilling or electron and hole doping.
}
\label{fig:j12cw}
\end{figure}

Figure \ref{fig:j12cw} depicts the band structure for SrCr$_2$As$_2$ and the contributions of $d$-orbitals to $J_1$ and $J_2$.
The band structure shown in the top panel highlights the changes in the bands upon doping. 
Since the band structure depicts only one path while the calculation of exchange coupling includes integral over the whole Brillouin zone, it is more informative to examine the orbital-resolved contributions (shown in the bottom two panels) to the exchange interaction. 
In the energy range of \SIrange{-0.25}{0.32}{\eV} near the Fermi level $\ef$, $d_{xy}$ and $d_{x^2-y^2}$ orbital-resolved $J_1$ values continue to increase and remain positive, indicating their contributions to AFM interactions. In contrast, the $d_{z^2}$ orbital-resolved values decrease and become negative with electron doping. The $d_{yz}$ and $d_{xz}$ orbital-resolved $J_1$ values are degenerate and remain negligible throughout this energy range.
For $J_{2}$, all orbital-resolved values are negative, suggesting a contribution to FM interactions.
Among these, the dominant $d_{x^2-y^2}$ orbital-resolved $J_2$ value continue to decrease while $d_{z^2}$ orbital-resolved value generally increases. 
Together, these trends govern the behavior of $J_2$ in this energy range.
These observations are consistent with the results shown in \rfig{fig:j12cd}.

For comparison, we also carried out the same calculations for $\bamnas$. 
The on-site Mn moment in $\bamnas$ is quite stable, around $\mu_\text{Mn}=3.6~\mu_\text{B}$ for different spin configurations observed in our DFT calculations. 
This contrasts with the significant variation in the Cr moment for different spin configurations in $\srcras$, indicating that total energy mapping is appropriate for $\bamnas$ while it becomes challenging for $\srcras$. 
Our linear response calculation yields $J_{2}/J_{1}=0.44$ in $\bamnas$, in good agreement with the experimental $J_{2}/J_{1}=0.33$~\cite{Johnston2011,Ramazanoglu2017}, suggesting a frustrated G-type magnetic ordering, in contrast with the stable G-type magnetic ordering in $\srcras$.
Orbital-resolved $J_2$ shows that its dominant contribution comes from the $d_{yz}$ and $d_{xz}$ 
orbitals in $\bamnas$, whereas the dominant contribution for $J_2$ in $\srcras$ comes from the $d_{x^2-y^2}$ orbital. 
Both $\bamnas$ and $\srcras$ have weak interlayer exchange couplings, with $J_c/J_1=0.15$ in $\bamnas$ and $J_c/J_1=0.05$ in $\srcras$, indicating their quasi-2D nature.
The N\'{e}el temperature is estimated within the mean-field approximation (MFA), yielding \SI{723}{\K} in $\srcras$ and \SI{826}{\K} in $\bamnas$, which are larger than the experimental values of 591~K \cite{Das2017} and 625~K \cite{Singh2009}, respectively. 
The overestimation of $T_{\text{C}}$ in MFA is well expected; especially for quasi-2D materials.
For true-2D materials, methods beyond MFA are desired to accurately estimate the critical temperature~\cite{mkhitaryan2021prb}.


\begin{figure}[htbp]
\includegraphics[width=1.00\linewidth,clip]{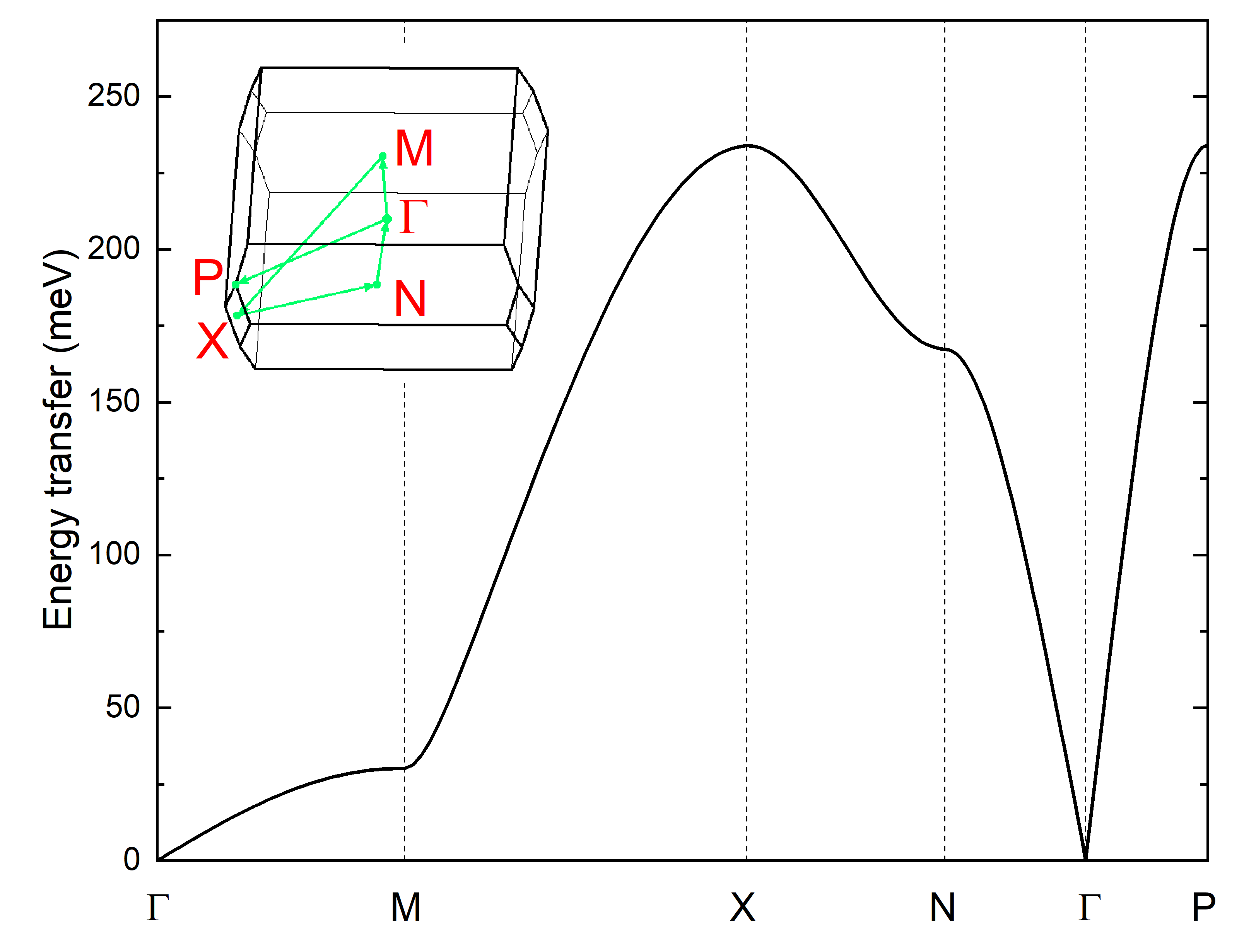}
\caption{Calculated spinwave spectrum of $\srcras$. 
The magnon band structure in $\srcras$ is calculated using the $J_1$-$J_{2}$-$J_c$ model with parameters obtained from DFT.
Magnetocrystalline anisotropy is not included.
The high-symmetry $k$ points $\Gamma$--M--X--N--$\Gamma$--P are illustrated in the inset.}
\label{Fig:sw}
\end{figure}

\begin{figure}
\includegraphics[scale=0.95]{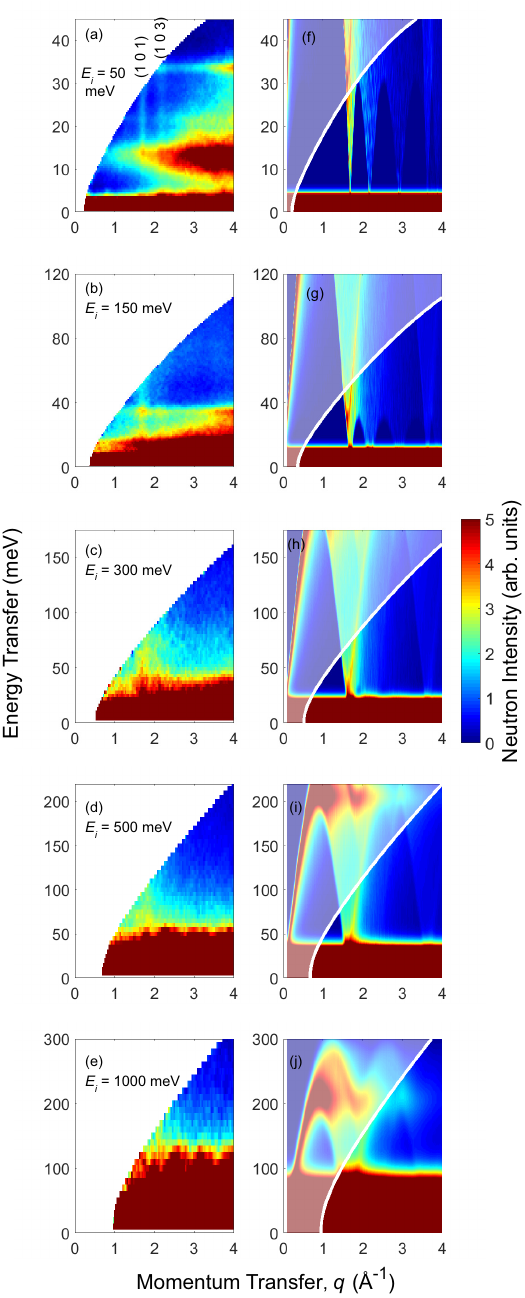}
\caption{(Color online) (a-e) Energy ($E$) versus momentum transfer ($q$) spectra of SrCr$_2$As$_2$ from the INS measurements with $E_i = 50, 150, 300, 500$ and $1000$ meV, respectively. 
(f-j) Calculated spin-wave excitations with the theoretical magnetic exchange parameters.
The white lines represent the experimental cut-offs. }
\label{Fig:Sqw}
\end{figure}

\subsection{Spin-wave spectra}\label{sec:INSresults}
Based on the G-type ground state and employing the predominant $J_1$ and $J_2$ values listed in \rtbl{tbl:j12_orb}, we perform magnon-dispersion calculations using the linear spin-wave theory (LSWT). 
The computed SW spectra along the high-symmetry directions $\Gamma$--M--X--N--$\Gamma$--P are illustrated in Fig.~\ref{Fig:sw}, with the spectrum reaching peaks as high as 234 meV.
It is important to note that $\Gamma$--X [$(-\xi,\xi,0)$] corresponds to the $J_{1}$ direction, while $\Gamma$--N [$(0,\xi,0)$] is along the $J_2$ direction. 
Subsequently, in the analysis of INS, we utilize the theoretical values of the exchange couplings presented in Table~\ref{tbl:j12_orb}.

We performed inelastic neutron-scattering experiments to test the validity of the first-principles calculations of the magnetic interactions.
To accurately isolate the experimental magnetic spectra (Fig.\ \ref{Fig:Sqw}), particularly in the low-$Q$ range where the magnetic form factor of Cr$^{2+}$ is significant, it is crucial to estimate and account for the background contributions. 
In our INS measurements, we consider two main background sources, Al phonons originating from the sample can and H impurities resulting from inadvertent moisture adsorption on the polycrystalline surfaces exposed to air.
The background contributions from Al phonons are measured separately by acquiring spectra with an empty Al can for each energy. 
These spectra are then scaled and subtracted from the measured data to remove the Al-related background.

Regarding the H scattering contribution, it becomes more pronounced in the high-$Q$ regime. 
This results in a broad feature with a $Q^2$ dependence that extends beyond 100 meV. 
To estimate the contribution from H scattering, we use a Gaussian function of the form $S_{\textrm{H}}(Q,E) = A_0 \exp[(E - E_r)^2/\Delta^2]$, where $E_r = \hbar^2Q^2/2m$ represents the recoil energy of H with a mass $m$. 
The parameters $\Delta$ and $A_0$ denote the energy width and scale factor, respectively.
After applying the corrections for both Al phonons and H scattering, we obtained the spectra as depicted in Figs.~\ref{Fig:Sqw}(a-e), which represent the experimental data with the background contributions subtracted.

\begin{figure}
\includegraphics[scale=1.0]{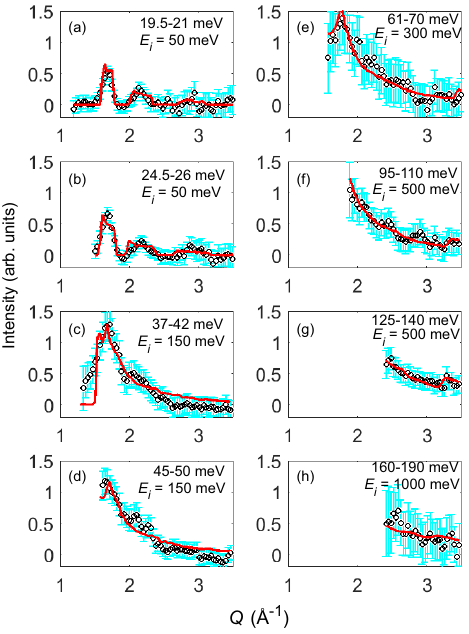}
\caption{(Color online) $Q$-cuts of the INS data at specified energy intervals are shown as black circles. The red lines are the simulated scattering intensities calculated using the magnetic exchange parameters obtained from the theoretical model.
Error bar magnitudes reflect statistical uncertainties from count rates and propagated errors due to background subtraction. Higher incident neutron energies ($E_i$), which are used to access broader $S({\bf Q}, E)$ ranges, affect count rates and consequently the size of the error bars.}
\label{Fig:Cuts1}
\end{figure}

The INS data in Figs.~\ref{Fig:Sqw}(a)-\ref{Fig:Sqw}(e) show the $Q$, $E$-dependent magnetic spectra at several different incident energies.  
For powder samples, the spectra are effectively averaged over all directions of $\bf Q$ and the neutron intensity $S(Q,E)$ is plotted.
At low incident energies shown in Figs.~\ref{Fig:Sqw} (a) and (b), the presence of strong phonon scattering from the sample leads to horizontal bands, with a maximum phonon energy of $\sim35$ meV, making the analysis of the magnetic spectrum challenging. 
However, even in the presence of these phonon contributions, we can still observe distinct vertical columns of magnetic excitations emerging from the magnetic (1 0 1) and (1 0 3) $\Gamma$ points, corresponding to $Q$ = 1.68 and 2.17 $\rm{\AA}^{-1}$, respectively. 
These magnetic excitations are observed to persist up to at least 175 meV as the incident energy $E_i$ is increased, as illustrated in Fig.~\ref{Fig:Cuts1}.
The large energy scale is shown in Fig.~\ref{Fig:Cuts1}(h), where the constant-energy $Q$-cut was done with a bin size of 160-190 meV (30 meV bin size for 3\% resolution at 1000 meV incident neutron energy, $E_i$, with a mean value of 175 meV), is in agreement with first-principles-based calculations shown in Fig.~\ref{Fig:sw}.

In Figs.~\ref{Fig:Sqw}(f)-\ref{Fig:Sqw}(j), we compare the INS data to SW calculations that are performed using the magnetic exchange parameters obtained from theoretical calculations ($J_1 = 25.1$ meV and $J_2 = -17.0$ meV), along with a weak effective interlayer coupling $J_{\rm{c}} = 1.2$ meV. 
The powder-averaging of $S({\bf Q}, E)$ is performed by Monte Carlo sampling of 25,000 $\textbf{Q}$ vectors for a given magnitude of $Q$ ranging from 0.1 to 4.0 \AA$^{-1}$, with a step size of 0.007 \AA$^{-1}$. 
The calculated neutron-scattering spectra, as illustrated in Figs. \ref{Fig:Sqw}(f-j) with the white lines representing the experimental kinematic limits, indicate good agreement between the model and the experimental measurements.

In Fig.~\ref{Fig:Cuts1}, we compare the calculated magnetic scattering intensities to INS data for several constant-energy Q-cuts.
The observed magnetic scattering appears weak compared to the background scattering, which has two major contributions.
First, there is a constant incoherent scattering close to $E = 0$. 
Second, there is a scattering due to phonons, which exhibits a $Q^2$ dependence. 
Modeling the complete phonon scattering without knowing all the vibrational modes of the system is challenging. However, it can be analytically represented as a background function $C_1 + C_2Q^2$, where $C_1$ and $C_2$ are fitted constants. 
The magnetic scattering is then extracted by subtracting both the incoherent and the phonon scattering, as illustrated in Fig.~\ref{Fig:Cuts1}.
To quantitatively compare the theoretical calculations with the experimental data, a series of $Q$-cuts at different energy transfers $E$ are performed, as depicted in Figs.~\ref{Fig:Cuts1}(a-h). 
The magnetic intensity in the $Q$-cuts is more pronounced and exhibits distinct features in the low-$Q$ regime. 
As $Q$ increases, the magnetic intensity gradually falls off in accordance with the Cr$^{2+}$ magnetic form factor. 
Overall, the experimental $Q$-cuts shown in Figs~\ref{Fig:Cuts1}(a-h) agree with the model calculations represented by the solid lines.

\section{Conclusion}

The experimental measurements and theoretical calculations within the LDA reveal the spin configurations, exchange parameters, and magnetic excitations in the $\srcras$ compound.
The stability of the $G$-type magnetic ordering in $\srcras$ is confirmed by both the total energy calculations and the presence of AFM $J_{\rm{1}}$ and FM $J_{\rm{2}}$ interactions, with $J_{\rm{2}}$/$J_{\rm{1}} \approx -0.68$. 
This indicates robust itinerant antiferromagnetism in $\srcras$ without magnetic frustration arising from competing interactions, distinguishing it from other TM arsenides based on Mn or Fe.

The dominant contribution to the NN AFM coupling $J_1$ originates from direct exchange mediated by the Cr $d_{xy}$ orbital, whereas the primary contribution to NNN FM $J_2$ arises from the $d_{x^2-y^2}$ and $d_{z^2}$ orbitals. 
The energy difference between $C$-type and $G$-type spin configurations is found to be small and dependent on the exchange functional used in the calculations.
This indicates a strong quasi-2D magnetism for SrCr$_2$As$_2$.

The estimated N\'{e}el temperature based on theoretical calculations, taking into account the $J_1$-$J_2$-$J_c$ Heisenberg model, is in good agreement with the experimental value.
The spin-wave spectrum calculated from the model displays magnetic excitations reaching energies as high as $234$ meV.
These results are consistent with experimental INS measurements for the spin excitations from powder samples.
We observe distinct magnetic excitations with steep dispersion originating from the magnetic $\Gamma$ points
and energy scales in good agreement with the theoretical calculations. 

Importantly, rigid band calculations suggest that carrier doping does not induce magnetic frustration in the system. The robust G-type antiferromagnetic order in SrCr$_2$As$_2$ implies that long-wavelength spin fluctuations may be relatively weak. This suggests that, under chemical substitution toward Fe-based superconductors, SrCr$_2$As$_2$ may be less likely to exhibit superconductivity driven by frustration-induced spin fluctuations, though other types of spin fluctuations may still play a role.
Notably, the N\'{e}el temperature remains nearly constant under small hole doping, a behavior analogous to that observed in BaMn$_2$As$_2$.

Overall, the concurrence of theoretical and experimental investigations indicate that first-principles calculations are able to establish the key energy scales while providing insight into the origin of the magnetic interactions in SrCr$_2$As$_2$.
This robust itinerant antiferromagnetism, devoid of frustration, underscores the compound's unique magnetic behavior and quasi-2D nature.

\acknowledgments

This research was supported by the U.S. Department of Energy, Office of Basic Energy Sciences, Division of Materials Sciences and Engineering.  Ames Laboratory is operated for the U.S. Department of Energy by Iowa State University under Contract No.~DE-AC02-07CH11358. A portion of this research used resources at the Spallation Neutron Source, a DOE Office of Science User Facility operated by the Oak Ridge National Laboratory.



\bibliography{srcras122.bib}

\begin{thebibliography}{51}%
\makeatletter
\providecommand \@ifxundefined [1]{%
 \@ifx{#1\undefined}
}%
\providecommand \@ifnum [1]{%
 \ifnum #1\expandafter \@firstoftwo
 \else \expandafter \@secondoftwo
 \fi
}%
\providecommand \@ifx [1]{%
 \ifx #1\expandafter \@firstoftwo
 \else \expandafter \@secondoftwo
 \fi
}%
\providecommand \natexlab [1]{#1}%
\providecommand \enquote  [1]{``#1''}%
\providecommand \bibnamefont  [1]{#1}%
\providecommand \bibfnamefont [1]{#1}%
\providecommand \citenamefont [1]{#1}%
\providecommand \href@noop [0]{\@secondoftwo}%
\providecommand \href [0]{\begingroup \@sanitize@url \@href}%
\providecommand \@href[1]{\@@startlink{#1}\@@href}%
\providecommand \@@href[1]{\endgroup#1\@@endlink}%
\providecommand \@sanitize@url [0]{\catcode `\\12\catcode `\$12\catcode
  `\&12\catcode `\#12\catcode `\^12\catcode `\_12\catcode `\%12\relax}%
\providecommand \@@startlink[1]{}%
\providecommand \@@endlink[0]{}%
\providecommand \url  [0]{\begingroup\@sanitize@url \@url }%
\providecommand \@url [1]{\endgroup\@href {#1}{\urlprefix }}%
\providecommand \urlprefix  [0]{URL }%
\providecommand \Eprint [0]{\href }%
\providecommand \doibase [0]{https://doi.org/}%
\providecommand \selectlanguage [0]{\@gobble}%
\providecommand \bibinfo  [0]{\@secondoftwo}%
\providecommand \bibfield  [0]{\@secondoftwo}%
\providecommand \translation [1]{[#1]}%
\providecommand \BibitemOpen [0]{}%
\providecommand \bibitemStop [0]{}%
\providecommand \bibitemNoStop [0]{.\EOS\space}%
\providecommand \EOS [0]{\spacefactor3000\relax}%
\providecommand \BibitemShut  [1]{\csname bibitem#1\endcsname}%
\let\auto@bib@innerbib\@empty
\bibitem [{\citenamefont {Johnston}(2010)}]{Johnston2010}%
  \BibitemOpen
  \bibfield  {author} {\bibinfo {author} {\bibfnamefont {D.~C.}\ \bibnamefont
  {Johnston}},\ }\bibfield  {title} {\bibinfo {title} {The puzzle of high
  temperature superconductivity in layered iron pnictides and chalcogenides},\
  }\href {https://doi.org/10.1080/00018732.2010.513480} {\bibfield  {journal}
  {\bibinfo  {journal} {Advances in Physics}\ }\textbf {\bibinfo {volume}
  {59}},\ \bibinfo {pages} {803} (\bibinfo {year} {2010})}\BibitemShut
  {NoStop}%
\bibitem [{\citenamefont {Dai}(2015)}]{Dai2015}%
  \BibitemOpen
  \bibfield  {author} {\bibinfo {author} {\bibfnamefont {P.}~\bibnamefont
  {Dai}},\ }\bibfield  {title} {\bibinfo {title} {Antiferromagnetic order and
  spin dynamics in iron-based superconductors},\ }\href
  {https://doi.org/10.1103/revmodphys.87.855} {\bibfield  {journal} {\bibinfo
  {journal} {Reviews of Modern Physics}\ }\textbf {\bibinfo {volume} {87}},\
  \bibinfo {pages} {855} (\bibinfo {year} {2015})}\BibitemShut {NoStop}%
\bibitem [{\citenamefont {Lumsden}\ and\ \citenamefont
  {Christianson}(2010)}]{Lumsden2010}%
  \BibitemOpen
  \bibfield  {author} {\bibinfo {author} {\bibfnamefont {M.~D.}\ \bibnamefont
  {Lumsden}}\ and\ \bibinfo {author} {\bibfnamefont {A.~D.}\ \bibnamefont
  {Christianson}},\ }\bibfield  {title} {\bibinfo {title} {{Magnetism in
  Fe-based superconductors}},\ }\href
  {https://doi.org/10.1088/0953-8984/22/20/203203} {\bibfield  {journal}
  {\bibinfo  {journal} {Journal of Physics: Condensed Matter}\ }\textbf
  {\bibinfo {volume} {22}},\ \bibinfo {pages} {203203} (\bibinfo {year}
  {2010})}\BibitemShut {NoStop}%
\bibitem [{\citenamefont {Inosov}(2016)}]{Inosov2016}%
  \BibitemOpen
  \bibfield  {author} {\bibinfo {author} {\bibfnamefont {D.~S.}\ \bibnamefont
  {Inosov}},\ }\bibfield  {title} {\bibinfo {title} {Spin fluctuations in iron
  pnictides and chalcogenides: From antiferromagnetism to superconductivity},\
  }\href {https://doi.org/https://doi.org/10.1016/j.crhy.2015.03.001}
  {\bibfield  {journal} {\bibinfo  {journal} {Comptes Rendus Physique}\
  }\textbf {\bibinfo {volume} {17}},\ \bibinfo {pages} {60} (\bibinfo {year}
  {2016})},\ \bibinfo {note} {iron-based superconductors / Supraconducteurs à
  base de fer}\BibitemShut {NoStop}%
\bibitem [{\citenamefont {Stewart}(2011)}]{Stewart2011}%
  \BibitemOpen
  \bibfield  {author} {\bibinfo {author} {\bibfnamefont {G.~R.}\ \bibnamefont
  {Stewart}},\ }\bibfield  {title} {\bibinfo {title} {Superconductivity in iron
  compounds},\ }\href {https://doi.org/10.1103/revmodphys.83.1589} {\bibfield
  {journal} {\bibinfo  {journal} {Reviews of Modern Physics}\ }\textbf
  {\bibinfo {volume} {83}},\ \bibinfo {pages} {1589} (\bibinfo {year}
  {2011})}\BibitemShut {NoStop}%
\bibitem [{\citenamefont {Si}\ \emph {et~al.}(2016)\citenamefont {Si},
  \citenamefont {Yu},\ and\ \citenamefont {Abrahams}}]{Si2016}%
  \BibitemOpen
  \bibfield  {author} {\bibinfo {author} {\bibfnamefont {Q.}~\bibnamefont
  {Si}}, \bibinfo {author} {\bibfnamefont {R.}~\bibnamefont {Yu}},\ and\
  \bibinfo {author} {\bibfnamefont {E.}~\bibnamefont {Abrahams}},\ }\bibfield
  {title} {\bibinfo {title} {High-temperature superconductivity in iron
  pnictides and chalcogenides},\ }\href@noop {} {\bibfield  {journal} {\bibinfo
   {journal} {Nature Reviews Materials}\ }\textbf {\bibinfo {volume} {1}}
  (\bibinfo {year} {2016})}\BibitemShut {NoStop}%
\bibitem [{\citenamefont {Scalapino}(2012)}]{Scalapino2012}%
  \BibitemOpen
  \bibfield  {author} {\bibinfo {author} {\bibfnamefont {D.~J.}\ \bibnamefont
  {Scalapino}},\ }\bibfield  {title} {\bibinfo {title} {A common thread: The
  pairing interaction for unconventional superconductors},\ }\href
  {https://doi.org/10.1103/revmodphys.84.1383} {\bibfield  {journal} {\bibinfo
  {journal} {Reviews of Modern Physics}\ }\textbf {\bibinfo {volume} {84}},\
  \bibinfo {pages} {1383} (\bibinfo {year} {2012})}\BibitemShut {NoStop}%
\bibitem [{\citenamefont {Vaknin}\ \emph {et~al.}(1987)\citenamefont {Vaknin},
  \citenamefont {Sinha}, \citenamefont {Moncton}, \citenamefont {Johnston},
  \citenamefont {Newsam}, \citenamefont {Safinya},\ and\ \citenamefont
  {King}}]{Vaknin1987}%
  \BibitemOpen
  \bibfield  {author} {\bibinfo {author} {\bibfnamefont {D.}~\bibnamefont
  {Vaknin}}, \bibinfo {author} {\bibfnamefont {S.~K.}\ \bibnamefont {Sinha}},
  \bibinfo {author} {\bibfnamefont {D.~E.}\ \bibnamefont {Moncton}}, \bibinfo
  {author} {\bibfnamefont {D.~C.}\ \bibnamefont {Johnston}}, \bibinfo {author}
  {\bibfnamefont {J.~M.}\ \bibnamefont {Newsam}}, \bibinfo {author}
  {\bibfnamefont {C.~R.}\ \bibnamefont {Safinya}},\ and\ \bibinfo {author}
  {\bibfnamefont {H.~E.}\ \bibnamefont {King}},\ }\bibfield  {title} {\bibinfo
  {title} {Antiferromagnetism in
  {La}$_{2}${CuO}$_{4\mathrm{\ensuremath{-}}\mathrm{y}}$},\ }\href
  {https://doi.org/10.1103/PhysRevLett.58.2802} {\bibfield  {journal} {\bibinfo
   {journal} {Phys. Rev. Lett.}\ }\textbf {\bibinfo {volume} {58}},\ \bibinfo
  {pages} {2802} (\bibinfo {year} {1987})}\BibitemShut {NoStop}%
\bibitem [{\citenamefont {Tranquada}\ \emph {et~al.}(1988)\citenamefont
  {Tranquada}, \citenamefont {Cox}, \citenamefont {Kunnmann}, \citenamefont
  {Moudden}, \citenamefont {Shirane}, \citenamefont {Suenaga}, \citenamefont
  {Zolliker}, \citenamefont {Vaknin}, \citenamefont {Sinha}, \citenamefont
  {Alvarez}, \citenamefont {Jacobson},\ and\ \citenamefont
  {Johnston}}]{Tranquada1988}%
  \BibitemOpen
  \bibfield  {author} {\bibinfo {author} {\bibfnamefont {J.~M.}\ \bibnamefont
  {Tranquada}}, \bibinfo {author} {\bibfnamefont {D.~E.}\ \bibnamefont {Cox}},
  \bibinfo {author} {\bibfnamefont {W.}~\bibnamefont {Kunnmann}}, \bibinfo
  {author} {\bibfnamefont {H.}~\bibnamefont {Moudden}}, \bibinfo {author}
  {\bibfnamefont {G.}~\bibnamefont {Shirane}}, \bibinfo {author} {\bibfnamefont
  {M.}~\bibnamefont {Suenaga}}, \bibinfo {author} {\bibfnamefont
  {P.}~\bibnamefont {Zolliker}}, \bibinfo {author} {\bibfnamefont
  {D.}~\bibnamefont {Vaknin}}, \bibinfo {author} {\bibfnamefont {S.~K.}\
  \bibnamefont {Sinha}}, \bibinfo {author} {\bibfnamefont {M.~S.}\ \bibnamefont
  {Alvarez}}, \bibinfo {author} {\bibfnamefont {A.~J.}\ \bibnamefont
  {Jacobson}},\ and\ \bibinfo {author} {\bibfnamefont {D.~C.}\ \bibnamefont
  {Johnston}},\ }\bibfield  {title} {\bibinfo {title} {{Neutron-Diffraction
  Determination of Antiferromagnetic Structure of {Cu} Ions in
  {YBa}$_{2}${Cu}$_{3}${O}$_{6+x}$ with $x=0.0$ and $0.15$}},\ }\href@noop {}
  {\bibfield  {journal} {\bibinfo  {journal} {Phys. Rev. Lett.}\ }\textbf
  {\bibinfo {volume} {60}},\ \bibinfo {pages} {156} (\bibinfo {year}
  {1988})}\BibitemShut {NoStop}%
\bibitem [{\citenamefont {Canfield}\ and\ \citenamefont
  {Bud'ko}(2010)}]{Canfield2010}%
  \BibitemOpen
  \bibfield  {author} {\bibinfo {author} {\bibfnamefont {P.~C.}\ \bibnamefont
  {Canfield}}\ and\ \bibinfo {author} {\bibfnamefont {S.~L.}\ \bibnamefont
  {Bud'ko}},\ }\bibfield  {title} {\bibinfo {title} {{FeAs-Based
  Superconductivity: A Case Study of the Effects of Transition Metal Doping on
  {BaFe}$_2${As}$_2$}},\ }\href@noop {} {\bibfield  {journal} {\bibinfo
  {journal} {Annual Review of Condensed Matter Physics}\ }\textbf {\bibinfo
  {volume} {1}},\ \bibinfo {pages} {27} (\bibinfo {year} {2010})}\BibitemShut
  {NoStop}%
\bibitem [{\citenamefont {Nedi\'{c}}\ \emph {et~al.}(2023)\citenamefont
  {Nedi\'{c}}, \citenamefont {Christensen}, \citenamefont {Lee}, \citenamefont
  {Li}, \citenamefont {Ueland}, \citenamefont {Fernandes}, \citenamefont
  {McQueeney}, \citenamefont {Ke},\ and\ \citenamefont {Orth}}]{nedic2023prb}%
  \BibitemOpen
  \bibfield  {author} {\bibinfo {author} {\bibfnamefont {A.-M.}\ \bibnamefont
  {Nedi\'{c}}}, \bibinfo {author} {\bibfnamefont {M.~H.}\ \bibnamefont
  {Christensen}}, \bibinfo {author} {\bibfnamefont {Y.}~\bibnamefont {Lee}},
  \bibinfo {author} {\bibfnamefont {B.}~\bibnamefont {Li}}, \bibinfo {author}
  {\bibfnamefont {B.~G.}\ \bibnamefont {Ueland}}, \bibinfo {author}
  {\bibfnamefont {R.~M.}\ \bibnamefont {Fernandes}}, \bibinfo {author}
  {\bibfnamefont {R.~J.}\ \bibnamefont {McQueeney}}, \bibinfo {author}
  {\bibfnamefont {L.}~\bibnamefont {Ke}},\ and\ \bibinfo {author}
  {\bibfnamefont {P.~P.}\ \bibnamefont {Orth}},\ }\bibfield  {title} {\bibinfo
  {title} {Competing magnetic fluctuations and orders in a multiorbital model
  of doped {SrCo$_2$As$_2$}},\ }\href
  {https://doi.org/10.1103/PhysRevB.108.245149} {\bibfield  {journal} {\bibinfo
   {journal} {Phys. Rev. B}\ }\textbf {\bibinfo {volume} {108}},\ \bibinfo
  {pages} {245149} (\bibinfo {year} {2023})}\BibitemShut {NoStop}%
\bibitem [{\citenamefont {Quirinale}\ \emph {et~al.}(2013)\citenamefont
  {Quirinale}, \citenamefont {Anand}, \citenamefont {Kim}, \citenamefont
  {Pandey}, \citenamefont {Huq}, \citenamefont {Stephens}, \citenamefont
  {Heitmann}, \citenamefont {Kreyssig}, \citenamefont {McQueeney},
  \citenamefont {Johnston},\ and\ \citenamefont {Goldman}}]{Quirinale2013}%
  \BibitemOpen
  \bibfield  {author} {\bibinfo {author} {\bibfnamefont {D.~G.}\ \bibnamefont
  {Quirinale}}, \bibinfo {author} {\bibfnamefont {V.~K.}\ \bibnamefont
  {Anand}}, \bibinfo {author} {\bibfnamefont {M.~G.}\ \bibnamefont {Kim}},
  \bibinfo {author} {\bibfnamefont {A.}~\bibnamefont {Pandey}}, \bibinfo
  {author} {\bibfnamefont {A.}~\bibnamefont {Huq}}, \bibinfo {author}
  {\bibfnamefont {P.~W.}\ \bibnamefont {Stephens}}, \bibinfo {author}
  {\bibfnamefont {T.~W.}\ \bibnamefont {Heitmann}}, \bibinfo {author}
  {\bibfnamefont {A.}~\bibnamefont {Kreyssig}}, \bibinfo {author}
  {\bibfnamefont {R.~J.}\ \bibnamefont {McQueeney}}, \bibinfo {author}
  {\bibfnamefont {D.~C.}\ \bibnamefont {Johnston}},\ and\ \bibinfo {author}
  {\bibfnamefont {A.~I.}\ \bibnamefont {Goldman}},\ }\bibfield  {title}
  {\bibinfo {title} {Crystal and magnetic structure of
  {CaCo}${}_{1.86}${As}${}_{2}$ studied by x-ray and neutron diffraction},\
  }\href@noop {} {\bibfield  {journal} {\bibinfo  {journal} {Phys. Rev. B}\
  }\textbf {\bibinfo {volume} {88}},\ \bibinfo {pages} {174420} (\bibinfo
  {year} {2013})}\BibitemShut {NoStop}%
\bibitem [{\citenamefont {Singh}\ \emph
  {et~al.}(2009{\natexlab{a}})\citenamefont {Singh}, \citenamefont {Green},
  \citenamefont {Huang}, \citenamefont {Kreyssig}, \citenamefont {McQueeney},
  \citenamefont {Johnston},\ and\ \citenamefont {Goldman}}]{Singh2009}%
  \BibitemOpen
  \bibfield  {author} {\bibinfo {author} {\bibfnamefont {Y.}~\bibnamefont
  {Singh}}, \bibinfo {author} {\bibfnamefont {M.~A.}\ \bibnamefont {Green}},
  \bibinfo {author} {\bibfnamefont {Q.}~\bibnamefont {Huang}}, \bibinfo
  {author} {\bibfnamefont {A.}~\bibnamefont {Kreyssig}}, \bibinfo {author}
  {\bibfnamefont {R.~J.}\ \bibnamefont {McQueeney}}, \bibinfo {author}
  {\bibfnamefont {D.~C.}\ \bibnamefont {Johnston}},\ and\ \bibinfo {author}
  {\bibfnamefont {A.~I.}\ \bibnamefont {Goldman}},\ }\bibfield  {title}
  {\bibinfo {title} {Magnetic order in {BaMn}$_{2}${As}$_{2}$ from neutron
  diffraction measurements},\ }\href
  {https://doi.org/10.1103/PhysRevB.80.100403} {\bibfield  {journal} {\bibinfo
  {journal} {Phys. Rev. B}\ }\textbf {\bibinfo {volume} {80}},\ \bibinfo
  {pages} {100403(R)} (\bibinfo {year} {2009}{\natexlab{a}})}\BibitemShut
  {NoStop}%
\bibitem [{\citenamefont {Shannon}\ \emph {et~al.}(2004)\citenamefont
  {Shannon}, \citenamefont {Schmidt}, \citenamefont {Penc},\ and\ \citenamefont
  {Thalmeier}}]{Shannon2004}%
  \BibitemOpen
  \bibfield  {author} {\bibinfo {author} {\bibfnamefont {N.}~\bibnamefont
  {Shannon}}, \bibinfo {author} {\bibfnamefont {B.}~\bibnamefont {Schmidt}},
  \bibinfo {author} {\bibfnamefont {K.}~\bibnamefont {Penc}},\ and\ \bibinfo
  {author} {\bibfnamefont {P.}~\bibnamefont {Thalmeier}},\ }\bibfield  {title}
  {\bibinfo {title} {Finite temperature properties and frustrated
  ferromagnetism in a square lattice {H}eisenberg model},\ }\href
  {https://doi.org/10.1140/epjb/e2004-00156-3} {\bibfield  {journal} {\bibinfo
  {journal} {The European Physical Journal B}\ }\textbf {\bibinfo {volume}
  {38}},\ \bibinfo {pages} {599} (\bibinfo {year} {2004})}\BibitemShut
  {NoStop}%
\bibitem [{\citenamefont {Sapkota}\ \emph {et~al.}(2017)\citenamefont
  {Sapkota}, \citenamefont {Ueland}, \citenamefont {Anand}, \citenamefont
  {Sangeetha}, \citenamefont {Abernathy}, \citenamefont {Stone}, \citenamefont
  {Niedziela}, \citenamefont {Johnston}, \citenamefont {Kreyssig},
  \citenamefont {Goldman},\ and\ \citenamefont {McQueeney}}]{Sapkota2017}%
  \BibitemOpen
  \bibfield  {author} {\bibinfo {author} {\bibfnamefont {A.}~\bibnamefont
  {Sapkota}}, \bibinfo {author} {\bibfnamefont {B.~G.}\ \bibnamefont {Ueland}},
  \bibinfo {author} {\bibfnamefont {V.~K.}\ \bibnamefont {Anand}}, \bibinfo
  {author} {\bibfnamefont {N.~S.}\ \bibnamefont {Sangeetha}}, \bibinfo {author}
  {\bibfnamefont {D.~L.}\ \bibnamefont {Abernathy}}, \bibinfo {author}
  {\bibfnamefont {M.~B.}\ \bibnamefont {Stone}}, \bibinfo {author}
  {\bibfnamefont {J.~L.}\ \bibnamefont {Niedziela}}, \bibinfo {author}
  {\bibfnamefont {D.~C.}\ \bibnamefont {Johnston}}, \bibinfo {author}
  {\bibfnamefont {A.}~\bibnamefont {Kreyssig}}, \bibinfo {author}
  {\bibfnamefont {A.~I.}\ \bibnamefont {Goldman}},\ and\ \bibinfo {author}
  {\bibfnamefont {R.~J.}\ \bibnamefont {McQueeney}},\ }\bibfield  {title}
  {\bibinfo {title} {Effective one-dimensional coupling in the highly
  frustrated square-lattice itinerant magnet
  {CaCo}$_{2\ensuremath{-}y}${As}$_{2}$},\ }\href
  {https://doi.org/10.1103/PhysRevLett.119.147201} {\bibfield  {journal}
  {\bibinfo  {journal} {Phys. Rev. Lett.}\ }\textbf {\bibinfo {volume} {119}},\
  \bibinfo {pages} {147201} (\bibinfo {year} {2017})}\BibitemShut {NoStop}%
\bibitem [{\citenamefont {Jayasekara}\ \emph {et~al.}(2013)\citenamefont
  {Jayasekara}, \citenamefont {Lee}, \citenamefont {Pandey}, \citenamefont
  {Tucker}, \citenamefont {Sapkota}, \citenamefont {Lamsal}, \citenamefont
  {Calder}, \citenamefont {Abernathy}, \citenamefont {Niedziela}, \citenamefont
  {Harmon}, \citenamefont {Kreyssig}, \citenamefont {Vaknin}, \citenamefont
  {Johnston}, \citenamefont {Goldman},\ and\ \citenamefont
  {McQueeney}}]{Jayasekara2013}%
  \BibitemOpen
  \bibfield  {author} {\bibinfo {author} {\bibfnamefont {W.}~\bibnamefont
  {Jayasekara}}, \bibinfo {author} {\bibfnamefont {Y.}~\bibnamefont {Lee}},
  \bibinfo {author} {\bibfnamefont {A.}~\bibnamefont {Pandey}}, \bibinfo
  {author} {\bibfnamefont {G.~S.}\ \bibnamefont {Tucker}}, \bibinfo {author}
  {\bibfnamefont {A.}~\bibnamefont {Sapkota}}, \bibinfo {author} {\bibfnamefont
  {J.}~\bibnamefont {Lamsal}}, \bibinfo {author} {\bibfnamefont
  {S.}~\bibnamefont {Calder}}, \bibinfo {author} {\bibfnamefont {D.~L.}\
  \bibnamefont {Abernathy}}, \bibinfo {author} {\bibfnamefont {J.~L.}\
  \bibnamefont {Niedziela}}, \bibinfo {author} {\bibfnamefont {B.~N.}\
  \bibnamefont {Harmon}}, \bibinfo {author} {\bibfnamefont {A.}~\bibnamefont
  {Kreyssig}}, \bibinfo {author} {\bibfnamefont {D.}~\bibnamefont {Vaknin}},
  \bibinfo {author} {\bibfnamefont {D.~C.}\ \bibnamefont {Johnston}}, \bibinfo
  {author} {\bibfnamefont {A.~I.}\ \bibnamefont {Goldman}},\ and\ \bibinfo
  {author} {\bibfnamefont {R.~J.}\ \bibnamefont {McQueeney}},\ }\bibfield
  {title} {\bibinfo {title} {Stripe antiferromagnetic spin fluctuations in
  ${\mathrm{srco}}_{2}{\mathrm{as}}_{2}$},\ }\href
  {https://doi.org/10.1103/PhysRevLett.111.157001} {\bibfield  {journal}
  {\bibinfo  {journal} {Phys. Rev. Lett.}\ }\textbf {\bibinfo {volume} {111}},\
  \bibinfo {pages} {157001} (\bibinfo {year} {2013})}\BibitemShut {NoStop}%
\bibitem [{\citenamefont {Chandra}\ \emph {et~al.}(1990)\citenamefont
  {Chandra}, \citenamefont {Coleman},\ and\ \citenamefont
  {Larkin}}]{Chandra1990}%
  \BibitemOpen
  \bibfield  {author} {\bibinfo {author} {\bibfnamefont {P.}~\bibnamefont
  {Chandra}}, \bibinfo {author} {\bibfnamefont {P.}~\bibnamefont {Coleman}},\
  and\ \bibinfo {author} {\bibfnamefont {A.~I.}\ \bibnamefont {Larkin}},\
  }\bibfield  {title} {\bibinfo {title} {{Ising Transition in Frustrated
  Heisenberg Models}},\ }\href@noop {} {\bibfield  {journal} {\bibinfo
  {journal} {Physical Review Letters}\ }\textbf {\bibinfo {volume} {64}},\
  \bibinfo {pages} {88} (\bibinfo {year} {1990})}\BibitemShut {NoStop}%
\bibitem [{\citenamefont {Shannon}\ \emph {et~al.}(2006)\citenamefont
  {Shannon}, \citenamefont {Momoi},\ and\ \citenamefont
  {Sindzingre}}]{Shannon2006}%
  \BibitemOpen
  \bibfield  {author} {\bibinfo {author} {\bibfnamefont {N.}~\bibnamefont
  {Shannon}}, \bibinfo {author} {\bibfnamefont {T.}~\bibnamefont {Momoi}},\
  and\ \bibinfo {author} {\bibfnamefont {P.}~\bibnamefont {Sindzingre}},\
  }\bibfield  {title} {\bibinfo {title} {{Nematic Order in Square Lattice
  Frustrated Ferromagnets}},\ }\href
  {https://doi.org/10.1103/physrevlett.96.027213} {\bibfield  {journal}
  {\bibinfo  {journal} {Physical Review Letters}\ }\textbf {\bibinfo {volume}
  {96}},\ \bibinfo {pages} {027213} (\bibinfo {year} {2006})}\BibitemShut
  {NoStop}%
\bibitem [{\citenamefont {Balents}(2010)}]{Balents2010}%
  \BibitemOpen
  \bibfield  {author} {\bibinfo {author} {\bibfnamefont {L.}~\bibnamefont
  {Balents}},\ }\bibfield  {title} {\bibinfo {title} {Spin liquids in
  frustrated magnets},\ }\href {https://doi.org/10.1038/nature08917} {\bibfield
   {journal} {\bibinfo  {journal} {Nature}\ }\textbf {\bibinfo {volume}
  {464}},\ \bibinfo {pages} {199} (\bibinfo {year} {2010})}\BibitemShut
  {NoStop}%
\bibitem [{\citenamefont {Savary}\ and\ \citenamefont
  {Balents}(2016)}]{Savary2016}%
  \BibitemOpen
  \bibfield  {author} {\bibinfo {author} {\bibfnamefont {L.}~\bibnamefont
  {Savary}}\ and\ \bibinfo {author} {\bibfnamefont {L.}~\bibnamefont
  {Balents}},\ }\bibfield  {title} {\bibinfo {title} {Quantum spin liquids: a
  review},\ }\href {https://doi.org/10.1088/0034-4885/80/1/016502} {\bibfield
  {journal} {\bibinfo  {journal} {Reports on Progress in Physics}\ }\textbf
  {\bibinfo {volume} {80}},\ \bibinfo {pages} {016502} (\bibinfo {year}
  {2016})}\BibitemShut {NoStop}%
\bibitem [{\citenamefont {Johnston}\ \emph {et~al.}(2011)\citenamefont
  {Johnston}, \citenamefont {McQueeney}, \citenamefont {Lake}, \citenamefont
  {Honecker}, \citenamefont {Zhitomirsky}, \citenamefont {Nath}, \citenamefont
  {Furukawa}, \citenamefont {Antropov},\ and\ \citenamefont
  {Singh}}]{Johnston2011}%
  \BibitemOpen
  \bibfield  {author} {\bibinfo {author} {\bibfnamefont {D.~C.}\ \bibnamefont
  {Johnston}}, \bibinfo {author} {\bibfnamefont {R.~J.}\ \bibnamefont
  {McQueeney}}, \bibinfo {author} {\bibfnamefont {B.}~\bibnamefont {Lake}},
  \bibinfo {author} {\bibfnamefont {A.}~\bibnamefont {Honecker}}, \bibinfo
  {author} {\bibfnamefont {M.~E.}\ \bibnamefont {Zhitomirsky}}, \bibinfo
  {author} {\bibfnamefont {R.}~\bibnamefont {Nath}}, \bibinfo {author}
  {\bibfnamefont {Y.}~\bibnamefont {Furukawa}}, \bibinfo {author}
  {\bibfnamefont {V.~P.}\ \bibnamefont {Antropov}},\ and\ \bibinfo {author}
  {\bibfnamefont {Y.}~\bibnamefont {Singh}},\ }\bibfield  {title} {\bibinfo
  {title} {Magnetic exchange interactions in {BaMn}$_2${As}$_2$: A case study
  of the {$J_1$-$J_2$-$J_c$ Heisenberg model}},\ }\href
  {https://doi.org/10.1103/physrevb.84.094445} {\bibfield  {journal} {\bibinfo
  {journal} {Physical Review B}\ }\textbf {\bibinfo {volume} {84}},\ \bibinfo
  {pages} {094445} (\bibinfo {year} {2011})}\BibitemShut {NoStop}%
\bibitem [{\citenamefont {Ramazanoglu}\ \emph {et~al.}(2017)\citenamefont
  {Ramazanoglu}, \citenamefont {Sapkota}, \citenamefont {Pandey}, \citenamefont
  {Lamsal}, \citenamefont {Abernathy}, \citenamefont {Niedziela}, \citenamefont
  {Stone}, \citenamefont {Kreyssig}, \citenamefont {Goldman}, \citenamefont
  {Johnston},\ and\ \citenamefont {McQueeney}}]{Ramazanoglu2017}%
  \BibitemOpen
  \bibfield  {author} {\bibinfo {author} {\bibfnamefont {M.}~\bibnamefont
  {Ramazanoglu}}, \bibinfo {author} {\bibfnamefont {A.}~\bibnamefont
  {Sapkota}}, \bibinfo {author} {\bibfnamefont {A.}~\bibnamefont {Pandey}},
  \bibinfo {author} {\bibfnamefont {J.}~\bibnamefont {Lamsal}}, \bibinfo
  {author} {\bibfnamefont {D.~L.}\ \bibnamefont {Abernathy}}, \bibinfo {author}
  {\bibfnamefont {J.~L.}\ \bibnamefont {Niedziela}}, \bibinfo {author}
  {\bibfnamefont {M.~B.}\ \bibnamefont {Stone}}, \bibinfo {author}
  {\bibfnamefont {A.}~\bibnamefont {Kreyssig}}, \bibinfo {author}
  {\bibfnamefont {A.~I.}\ \bibnamefont {Goldman}}, \bibinfo {author}
  {\bibfnamefont {D.~C.}\ \bibnamefont {Johnston}},\ and\ \bibinfo {author}
  {\bibfnamefont {R.~J.}\ \bibnamefont {McQueeney}},\ }\bibfield  {title}
  {\bibinfo {title} {Robust antiferromagnetic spin waves across the
  metal-insulator transition in hole-doped {BaMn}$_{2}${As}$_{2}$},\ }\href
  {https://doi.org/10.1103/PhysRevB.95.224401} {\bibfield  {journal} {\bibinfo
  {journal} {Phys. Rev. B}\ }\textbf {\bibinfo {volume} {95}},\ \bibinfo
  {pages} {224401} (\bibinfo {year} {2017})}\BibitemShut {NoStop}%
\bibitem [{\citenamefont {Das}\ \emph {et~al.}(2017)\citenamefont {Das},
  \citenamefont {Sangeetha}, \citenamefont {Lindemann}, \citenamefont
  {Heitmann}, \citenamefont {Kreyssig}, \citenamefont {Goldman}, \citenamefont
  {McQueeney}, \citenamefont {Johnston},\ and\ \citenamefont
  {Vaknin}}]{Das2017}%
  \BibitemOpen
  \bibfield  {author} {\bibinfo {author} {\bibfnamefont {P.}~\bibnamefont
  {Das}}, \bibinfo {author} {\bibfnamefont {N.~S.}\ \bibnamefont {Sangeetha}},
  \bibinfo {author} {\bibfnamefont {G.~R.}\ \bibnamefont {Lindemann}}, \bibinfo
  {author} {\bibfnamefont {T.~W.}\ \bibnamefont {Heitmann}}, \bibinfo {author}
  {\bibfnamefont {A.}~\bibnamefont {Kreyssig}}, \bibinfo {author}
  {\bibfnamefont {A.~I.}\ \bibnamefont {Goldman}}, \bibinfo {author}
  {\bibfnamefont {R.~J.}\ \bibnamefont {McQueeney}}, \bibinfo {author}
  {\bibfnamefont {D.~C.}\ \bibnamefont {Johnston}},\ and\ \bibinfo {author}
  {\bibfnamefont {D.}~\bibnamefont {Vaknin}},\ }\bibfield  {title} {\bibinfo
  {title} {{Itinerant G-type antiferromagnetic order in {SrCr}$_2${As}$_2$}},\
  }\href {https://doi.org/10.1103/physrevb.96.014411} {\bibfield  {journal}
  {\bibinfo  {journal} {Physical Review B}\ }\textbf {\bibinfo {volume} {96}},\
  \bibinfo {pages} {014411} (\bibinfo {year} {2017})}\BibitemShut {NoStop}%
\bibitem [{\citenamefont {An}\ \emph {et~al.}(2009)\citenamefont {An},
  \citenamefont {Sefat}, \citenamefont {Singh},\ and\ \citenamefont
  {Du}}]{An2009}%
  \BibitemOpen
  \bibfield  {author} {\bibinfo {author} {\bibfnamefont {J.}~\bibnamefont
  {An}}, \bibinfo {author} {\bibfnamefont {A.~S.}\ \bibnamefont {Sefat}},
  \bibinfo {author} {\bibfnamefont {D.~J.}\ \bibnamefont {Singh}},\ and\
  \bibinfo {author} {\bibfnamefont {M.-H.}\ \bibnamefont {Du}},\ }\bibfield
  {title} {\bibinfo {title} {Electronic structure and magnetism in
  {BaMn}$_{2}${As}$_{2}$ and {BaMn}$_{2}${Sb}$_{2}$},\ }\href
  {https://doi.org/10.1103/PhysRevB.79.075120} {\bibfield  {journal} {\bibinfo
  {journal} {Phys. Rev. B}\ }\textbf {\bibinfo {volume} {79}},\ \bibinfo
  {pages} {075120} (\bibinfo {year} {2009})}\BibitemShut {NoStop}%
\bibitem [{\citenamefont {Singh}\ \emph
  {et~al.}(2009{\natexlab{b}})\citenamefont {Singh}, \citenamefont {Ellern},\
  and\ \citenamefont {Johnston}}]{Singh2009b}%
  \BibitemOpen
  \bibfield  {author} {\bibinfo {author} {\bibfnamefont {Y.}~\bibnamefont
  {Singh}}, \bibinfo {author} {\bibfnamefont {A.}~\bibnamefont {Ellern}},\ and\
  \bibinfo {author} {\bibfnamefont {D.~C.}\ \bibnamefont {Johnston}},\
  }\bibfield  {title} {\bibinfo {title} {Magnetic, transport, and thermal
  properties of single crystals of the layered arsenide
  {BaMn}$_{2}${As}$_{2}$},\ }\href {https://doi.org/10.1103/PhysRevB.79.094519}
  {\bibfield  {journal} {\bibinfo  {journal} {Phys. Rev. B}\ }\textbf {\bibinfo
  {volume} {79}},\ \bibinfo {pages} {094519} (\bibinfo {year}
  {2009}{\natexlab{b}})}\BibitemShut {NoStop}%
\bibitem [{\citenamefont {Singh}\ \emph
  {et~al.}(2009{\natexlab{c}})\citenamefont {Singh}, \citenamefont {Sefat},
  \citenamefont {McGuire}, \citenamefont {Sales}, \citenamefont {Mandrus},
  \citenamefont {VanBebber},\ and\ \citenamefont {Keppens}}]{SinghDJ2009}%
  \BibitemOpen
  \bibfield  {author} {\bibinfo {author} {\bibfnamefont {D.~J.}\ \bibnamefont
  {Singh}}, \bibinfo {author} {\bibfnamefont {A.~S.}\ \bibnamefont {Sefat}},
  \bibinfo {author} {\bibfnamefont {M.~A.}\ \bibnamefont {McGuire}}, \bibinfo
  {author} {\bibfnamefont {B.~C.}\ \bibnamefont {Sales}}, \bibinfo {author}
  {\bibfnamefont {D.}~\bibnamefont {Mandrus}}, \bibinfo {author} {\bibfnamefont
  {L.~H.}\ \bibnamefont {VanBebber}},\ and\ \bibinfo {author} {\bibfnamefont
  {V.}~\bibnamefont {Keppens}},\ }\bibfield  {title} {\bibinfo {title}
  {Itinerant antiferromagnetism in {BaCr}$_{2}${As}$_{2}$: Experimental
  characterization and electronic structure calculations},\ }\href
  {https://doi.org/10.1103/physrevb.79.094429} {\bibfield  {journal} {\bibinfo
  {journal} {Physical Review B}\ }\textbf {\bibinfo {volume} {79}},\ \bibinfo
  {pages} {094429} (\bibinfo {year} {2009}{\natexlab{c}})}\BibitemShut
  {NoStop}%
\bibitem [{\citenamefont {McNally}\ \emph {et~al.}(2015)\citenamefont
  {McNally}, \citenamefont {Zellman}, \citenamefont {Yin}, \citenamefont
  {Post}, \citenamefont {He}, \citenamefont {Hao}, \citenamefont {Kotliar},
  \citenamefont {Basov}, \citenamefont {Homes},\ and\ \citenamefont
  {Aronson}}]{McNally2015}%
  \BibitemOpen
  \bibfield  {author} {\bibinfo {author} {\bibfnamefont {D.~E.}\ \bibnamefont
  {McNally}}, \bibinfo {author} {\bibfnamefont {S.}~\bibnamefont {Zellman}},
  \bibinfo {author} {\bibfnamefont {Z.~P.}\ \bibnamefont {Yin}}, \bibinfo
  {author} {\bibfnamefont {K.~W.}\ \bibnamefont {Post}}, \bibinfo {author}
  {\bibfnamefont {H.}~\bibnamefont {He}}, \bibinfo {author} {\bibfnamefont
  {K.}~\bibnamefont {Hao}}, \bibinfo {author} {\bibfnamefont {G.}~\bibnamefont
  {Kotliar}}, \bibinfo {author} {\bibfnamefont {D.}~\bibnamefont {Basov}},
  \bibinfo {author} {\bibfnamefont {C.~C.}\ \bibnamefont {Homes}},\ and\
  \bibinfo {author} {\bibfnamefont {M.~C.}\ \bibnamefont {Aronson}},\
  }\bibfield  {title} {\bibinfo {title} {From hund's insulator to fermi liquid:
  Optical spectroscopy study of {K} doping in {BaMn}$_{2}${As}$_{2}$},\ }\href
  {https://doi.org/10.1103/physrevb.92.115142} {\bibfield  {journal} {\bibinfo
  {journal} {Physical Review B}\ }\textbf {\bibinfo {volume} {92}},\ \bibinfo
  {pages} {115142} (\bibinfo {year} {2015})}\BibitemShut {NoStop}%
\bibitem [{\citenamefont {Blaha}\ \emph {et~al.}(2020)\citenamefont {Blaha},
  \citenamefont {Schwarz}, \citenamefont {Tran}, \citenamefont {Laskowski},
  \citenamefont {Madsen},\ and\ \citenamefont {Marks}}]{Blaha2020}%
  \BibitemOpen
  \bibfield  {author} {\bibinfo {author} {\bibfnamefont {P.}~\bibnamefont
  {Blaha}}, \bibinfo {author} {\bibfnamefont {K.}~\bibnamefont {Schwarz}},
  \bibinfo {author} {\bibfnamefont {F.}~\bibnamefont {Tran}}, \bibinfo {author}
  {\bibfnamefont {R.}~\bibnamefont {Laskowski}}, \bibinfo {author}
  {\bibfnamefont {G.~K.~H.}\ \bibnamefont {Madsen}},\ and\ \bibinfo {author}
  {\bibfnamefont {L.~D.}\ \bibnamefont {Marks}},\ }\bibfield  {title} {\bibinfo
  {title} {{WIEN2k: An APW+lo program for calculating the properties of
  solids}},\ }\href@noop {} {\bibfield  {journal} {\bibinfo  {journal} {The
  Journal of Chemical Physics}\ }\textbf {\bibinfo {volume} {152}} (\bibinfo
  {year} {2020})},\ \bibinfo {note} {074101}\BibitemShut {NoStop}%
\bibitem [{\citenamefont {Blaha}\ \emph {et~al.}(2023)\citenamefont {Blaha},
  \citenamefont {Schwarz}, \citenamefont {Madsen}, \citenamefont {Kvasnicka},
  \citenamefont {Luitz}, \citenamefont {Laskowski}, \citenamefont {Tran},\ and\
  \citenamefont {Marks}}]{wien2k}%
  \BibitemOpen
  \bibfield  {author} {\bibinfo {author} {\bibfnamefont {P.}~\bibnamefont
  {Blaha}}, \bibinfo {author} {\bibfnamefont {K.}~\bibnamefont {Schwarz}},
  \bibinfo {author} {\bibfnamefont {G.~K.~H.}\ \bibnamefont {Madsen}}, \bibinfo
  {author} {\bibfnamefont {D.}~\bibnamefont {Kvasnicka}}, \bibinfo {author}
  {\bibfnamefont {J.}~\bibnamefont {Luitz}}, \bibinfo {author} {\bibfnamefont
  {R.}~\bibnamefont {Laskowski}}, \bibinfo {author} {\bibfnamefont
  {F.}~\bibnamefont {Tran}},\ and\ \bibinfo {author} {\bibfnamefont {L.~D.}\
  \bibnamefont {Marks}},\ }\href@noop {} {\emph {\bibinfo {title} {WIEN2k: An
  Augmented Plane Wave plus Local Orbitals Program for Calculating Crystal
  Properties}}}\ (\bibinfo  {publisher} {Vienna University of Technology},\
  \bibinfo {address} {Austria},\ \bibinfo {year} {2023})\BibitemShut {NoStop}%
\bibitem [{\citenamefont {von Barth}\ and\ \citenamefont
  {Hedin}(1972)}]{Barth1972}%
  \BibitemOpen
  \bibfield  {author} {\bibinfo {author} {\bibfnamefont {U.}~\bibnamefont {von
  Barth}}\ and\ \bibinfo {author} {\bibfnamefont {L.}~\bibnamefont {Hedin}},\
  }\bibfield  {title} {\bibinfo {title} {A local exchange-correlation potential
  for the spin polarized case. i},\ }\href@noop {} {\bibfield  {journal}
  {\bibinfo  {journal} {Journal of Physics C: Solid State Physics}\ }\textbf
  {\bibinfo {volume} {5}},\ \bibinfo {pages} {1629} (\bibinfo {year}
  {1972})}\BibitemShut {NoStop}%
\bibitem [{\citenamefont {Perdew}\ \emph {et~al.}(1996)\citenamefont {Perdew},
  \citenamefont {Burke},\ and\ \citenamefont {Ernzerhof}}]{Perdew1996}%
  \BibitemOpen
  \bibfield  {author} {\bibinfo {author} {\bibfnamefont {J.~P.}\ \bibnamefont
  {Perdew}}, \bibinfo {author} {\bibfnamefont {K.}~\bibnamefont {Burke}},\ and\
  \bibinfo {author} {\bibfnamefont {M.}~\bibnamefont {Ernzerhof}},\ }\bibfield
  {title} {\bibinfo {title} {{Generalized Gradient Approximation Made
  Simple}},\ }\href {https://doi.org/10.1103/PhysRevLett.77.3865} {\bibfield
  {journal} {\bibinfo  {journal} {Phys. Rev. Lett.}\ }\textbf {\bibinfo
  {volume} {77}},\ \bibinfo {pages} {3865} (\bibinfo {year}
  {1996})}\BibitemShut {NoStop}%
\bibitem [{\citenamefont {Marzari}\ and\ \citenamefont
  {Vanderbilt}(1997)}]{marzari1997prb}%
  \BibitemOpen
  \bibfield  {author} {\bibinfo {author} {\bibfnamefont {N.}~\bibnamefont
  {Marzari}}\ and\ \bibinfo {author} {\bibfnamefont {D.}~\bibnamefont
  {Vanderbilt}},\ }\bibfield  {title} {\bibinfo {title} {Maximally localized
  generalized wannier functions for composite energy bands},\ }\href
  {https://doi.org/10.1103/physrevb.56.12847} {\bibfield  {journal} {\bibinfo
  {journal} {Physical Review B}\ }\textbf {\bibinfo {volume} {56}},\ \bibinfo
  {pages} {12847} (\bibinfo {year} {1997})}\BibitemShut {NoStop}%
\bibitem [{\citenamefont {Souza}\ \emph {et~al.}(2001)\citenamefont {Souza},
  \citenamefont {Marzari},\ and\ \citenamefont {Vanderbilt}}]{souza2001prb}%
  \BibitemOpen
  \bibfield  {author} {\bibinfo {author} {\bibfnamefont {I.}~\bibnamefont
  {Souza}}, \bibinfo {author} {\bibfnamefont {N.}~\bibnamefont {Marzari}},\
  and\ \bibinfo {author} {\bibfnamefont {D.}~\bibnamefont {Vanderbilt}},\
  }\bibfield  {title} {\bibinfo {title} {Maximally localized wannier functions
  for entangled energy bands},\ }\href
  {https://doi.org/10.1103/PhysRevB.65.035109} {\bibfield  {journal} {\bibinfo
  {journal} {Phys. Rev. B}\ }\textbf {\bibinfo {volume} {65}},\ \bibinfo
  {pages} {035109} (\bibinfo {year} {2001})}\BibitemShut {NoStop}%
\bibitem [{\citenamefont {Marzari}\ \emph {et~al.}(2012)\citenamefont
  {Marzari}, \citenamefont {Mostofi}, \citenamefont {Yates}, \citenamefont
  {Souza},\ and\ \citenamefont {Vanderbilt}}]{marzari2012rmp}%
  \BibitemOpen
  \bibfield  {author} {\bibinfo {author} {\bibfnamefont {N.}~\bibnamefont
  {Marzari}}, \bibinfo {author} {\bibfnamefont {A.~A.}\ \bibnamefont
  {Mostofi}}, \bibinfo {author} {\bibfnamefont {J.~R.}\ \bibnamefont {Yates}},
  \bibinfo {author} {\bibfnamefont {I.}~\bibnamefont {Souza}},\ and\ \bibinfo
  {author} {\bibfnamefont {D.}~\bibnamefont {Vanderbilt}},\ }\bibfield  {title}
  {\bibinfo {title} {Maximally localized wannier functions: Theory and
  applications},\ }\href {https://doi.org/10.1103/RevModPhys.84.1419}
  {\bibfield  {journal} {\bibinfo  {journal} {Rev. Mod. Phys.}\ }\textbf
  {\bibinfo {volume} {84}},\ \bibinfo {pages} {1419} (\bibinfo {year}
  {2012})}\BibitemShut {NoStop}%
\bibitem [{\citenamefont {Mostofi}\ \emph {et~al.}(2014)\citenamefont
  {Mostofi}, \citenamefont {Yates}, \citenamefont {Pizzi}, \citenamefont {Lee},
  \citenamefont {Souza}, \citenamefont {Vanderbilt},\ and\ \citenamefont
  {Marzari}}]{mostofi2014cpc}%
  \BibitemOpen
  \bibfield  {author} {\bibinfo {author} {\bibfnamefont {A.~A.}\ \bibnamefont
  {Mostofi}}, \bibinfo {author} {\bibfnamefont {J.~R.}\ \bibnamefont {Yates}},
  \bibinfo {author} {\bibfnamefont {G.}~\bibnamefont {Pizzi}}, \bibinfo
  {author} {\bibfnamefont {Y.-S.}\ \bibnamefont {Lee}}, \bibinfo {author}
  {\bibfnamefont {I.}~\bibnamefont {Souza}}, \bibinfo {author} {\bibfnamefont
  {D.}~\bibnamefont {Vanderbilt}},\ and\ \bibinfo {author} {\bibfnamefont
  {N.}~\bibnamefont {Marzari}},\ }\bibfield  {title} {\bibinfo {title} {{An
  updated version of wannier90: A tool for obtaining maximally-localised
  Wannier functions}},\ }\href
  {https://doi.org/http://doi.org/10.1016/j.cpc.2014.05.003} {\bibfield
  {journal} {\bibinfo  {journal} {Computer Physics Communications}\ }\textbf
  {\bibinfo {volume} {185}},\ \bibinfo {pages} {2309} (\bibinfo {year}
  {2014})}\BibitemShut {NoStop}%
\bibitem [{\citenamefont {Liechtenstein}\ \emph {et~al.}(1984)\citenamefont
  {Liechtenstein}, \citenamefont {Katsnelson},\ and\ \citenamefont
  {Gubanov}}]{Liechtenstein1984}%
  \BibitemOpen
  \bibfield  {author} {\bibinfo {author} {\bibfnamefont {A.~I.}\ \bibnamefont
  {Liechtenstein}}, \bibinfo {author} {\bibfnamefont {M.~I.}\ \bibnamefont
  {Katsnelson}},\ and\ \bibinfo {author} {\bibfnamefont {V.~A.}\ \bibnamefont
  {Gubanov}},\ }\bibfield  {title} {\bibinfo {title} {Exchange interactions and
  spin-wave stiffness in ferromagnetic metals},\ }\href
  {https://doi.org/10.1088/0305-4608/14/7/007} {\bibfield  {journal} {\bibinfo
  {journal} {Journal of Physics F: Metal Physics}\ }\textbf {\bibinfo {volume}
  {14}},\ \bibinfo {pages} {L125} (\bibinfo {year} {1984})}\BibitemShut
  {NoStop}%
\bibitem [{\citenamefont {Rosenberg}\ \emph {et~al.}(2022)\citenamefont
  {Rosenberg}, \citenamefont {DeStefano}, \citenamefont {Guo}, \citenamefont
  {Oh}, \citenamefont {Hashimoto}, \citenamefont {Lu}, \citenamefont
  {Birgeneau}, \citenamefont {Lee}, \citenamefont {Ke}, \citenamefont {Yi},\
  and\ \citenamefont {Chu}}]{rosenberg2022prb}%
  \BibitemOpen
  \bibfield  {author} {\bibinfo {author} {\bibfnamefont {E.}~\bibnamefont
  {Rosenberg}}, \bibinfo {author} {\bibfnamefont {J.~M.}\ \bibnamefont
  {DeStefano}}, \bibinfo {author} {\bibfnamefont {Y.}~\bibnamefont {Guo}},
  \bibinfo {author} {\bibfnamefont {J.~S.}\ \bibnamefont {Oh}}, \bibinfo
  {author} {\bibfnamefont {M.}~\bibnamefont {Hashimoto}}, \bibinfo {author}
  {\bibfnamefont {D.}~\bibnamefont {Lu}}, \bibinfo {author} {\bibfnamefont
  {R.~J.}\ \bibnamefont {Birgeneau}}, \bibinfo {author} {\bibfnamefont
  {Y.}~\bibnamefont {Lee}}, \bibinfo {author} {\bibfnamefont {L.}~\bibnamefont
  {Ke}}, \bibinfo {author} {\bibfnamefont {M.}~\bibnamefont {Yi}},\ and\
  \bibinfo {author} {\bibfnamefont {J.-H.}\ \bibnamefont {Chu}},\ }\bibfield
  {title} {\bibinfo {title} {Uniaxial ferromagnetism in the kagome metal
  {TbV$_{6}$Sn$_6$}},\ }\href@noop {} {\bibfield  {journal} {\bibinfo
  {journal} {Phys. Rev. B}\ }\textbf {\bibinfo {volume} {106}},\ \bibinfo
  {pages} {115139} (\bibinfo {year} {2022})}\BibitemShut {NoStop}%
\bibitem [{\citenamefont {Lee}\ \emph {et~al.}(2023)\citenamefont {Lee},
  \citenamefont {Skomski}, \citenamefont {Wang}, \citenamefont {Orth},
  \citenamefont {Ren}, \citenamefont {Kang}, \citenamefont {Pathak},
  \citenamefont {Kutepov}, \citenamefont {Harmon}, \citenamefont {McQueeney},
  \citenamefont {Mazin},\ and\ \citenamefont {Ke}}]{lee2023prb}%
  \BibitemOpen
  \bibfield  {author} {\bibinfo {author} {\bibfnamefont {Y.}~\bibnamefont
  {Lee}}, \bibinfo {author} {\bibfnamefont {R.}~\bibnamefont {Skomski}},
  \bibinfo {author} {\bibfnamefont {X.}~\bibnamefont {Wang}}, \bibinfo {author}
  {\bibfnamefont {P.~P.}\ \bibnamefont {Orth}}, \bibinfo {author}
  {\bibfnamefont {Y.}~\bibnamefont {Ren}}, \bibinfo {author} {\bibfnamefont
  {B.}~\bibnamefont {Kang}}, \bibinfo {author} {\bibfnamefont {A.~K.}\
  \bibnamefont {Pathak}}, \bibinfo {author} {\bibfnamefont {A.}~\bibnamefont
  {Kutepov}}, \bibinfo {author} {\bibfnamefont {B.~N.}\ \bibnamefont {Harmon}},
  \bibinfo {author} {\bibfnamefont {R.~J.}\ \bibnamefont {McQueeney}}, \bibinfo
  {author} {\bibfnamefont {I.~I.}\ \bibnamefont {Mazin}},\ and\ \bibinfo
  {author} {\bibfnamefont {L.}~\bibnamefont {Ke}},\ }\bibfield  {title}
  {\bibinfo {title} {Interplay between magnetism and band topology in the
  kagome magnets {$R$Mn$_6$Sn$_6$}},\ }\href
  {https://doi.org/10.1103/PhysRevB.108.045132} {\bibfield  {journal} {\bibinfo
   {journal} {Phys. Rev. B}\ }\textbf {\bibinfo {volume} {108}},\ \bibinfo
  {pages} {045132} (\bibinfo {year} {2023})}\BibitemShut {NoStop}%
\bibitem [{\citenamefont {Timmons}\ \emph {et~al.}(2020)\citenamefont
  {Timmons}, \citenamefont {Teknowijoyo}, \citenamefont {Ko\'{n}czykowski},
  \citenamefont {Cavani}, \citenamefont {Tanatar}, \citenamefont {Ghimire},
  \citenamefont {Cho}, \citenamefont {Lee}, \citenamefont {Ke}, \citenamefont
  {Jo}, \citenamefont {Bud'ko}, \citenamefont {Canfield}, \citenamefont {Orth},
  \citenamefont {Scheurer},\ and\ \citenamefont {Prozorov}}]{timmons2020prr}%
  \BibitemOpen
  \bibfield  {author} {\bibinfo {author} {\bibfnamefont {E.~I.}\ \bibnamefont
  {Timmons}}, \bibinfo {author} {\bibfnamefont {S.}~\bibnamefont
  {Teknowijoyo}}, \bibinfo {author} {\bibfnamefont {M.}~\bibnamefont
  {Ko\'{n}czykowski}}, \bibinfo {author} {\bibfnamefont {O.}~\bibnamefont
  {Cavani}}, \bibinfo {author} {\bibfnamefont {M.~A.}\ \bibnamefont {Tanatar}},
  \bibinfo {author} {\bibfnamefont {S.}~\bibnamefont {Ghimire}}, \bibinfo
  {author} {\bibfnamefont {K.}~\bibnamefont {Cho}}, \bibinfo {author}
  {\bibfnamefont {Y.}~\bibnamefont {Lee}}, \bibinfo {author} {\bibfnamefont
  {L.}~\bibnamefont {Ke}}, \bibinfo {author} {\bibfnamefont {N.~H.}\
  \bibnamefont {Jo}}, \bibinfo {author} {\bibfnamefont {S.~L.}\ \bibnamefont
  {Bud'ko}}, \bibinfo {author} {\bibfnamefont {P.~C.}\ \bibnamefont
  {Canfield}}, \bibinfo {author} {\bibfnamefont {P.~P.}\ \bibnamefont {Orth}},
  \bibinfo {author} {\bibfnamefont {M.~S.}\ \bibnamefont {Scheurer}},\ and\
  \bibinfo {author} {\bibfnamefont {R.}~\bibnamefont {Prozorov}},\ }\bibfield
  {title} {\bibinfo {title} {{Electron irradiation effects on superconductivity
  in ${\mathrm{PdTe}}_{2}$: An application of a generalized Anderson
  theorem}},\ }\href@noop {} {\bibfield  {journal} {\bibinfo  {journal} {Phys.
  Rev. Research}\ }\textbf {\bibinfo {volume} {2}},\ \bibinfo {pages} {023140}
  (\bibinfo {year} {2020})}\BibitemShut {NoStop}%
\bibitem [{\citenamefont {Ke}(2019)}]{ke2019prb}%
  \BibitemOpen
  \bibfield  {author} {\bibinfo {author} {\bibfnamefont {L.}~\bibnamefont
  {Ke}},\ }\bibfield  {title} {\bibinfo {title} {Intersublattice
  magnetocrystalline anisotropy using a realistic tight-binding method based on
  maximally localized {Wannier} functions},\ }\href
  {https://doi.org/10.1103/PhysRevB.99.054418} {\bibfield  {journal} {\bibinfo
  {journal} {Phys. Rev. B}\ }\textbf {\bibinfo {volume} {99}},\ \bibinfo
  {pages} {054418} (\bibinfo {year} {2019})}\BibitemShut {NoStop}%
\bibitem [{\citenamefont {Lukashev}\ \emph {et~al.}(2023)\citenamefont
  {Lukashev}, \citenamefont {Ramker}, \citenamefont {Schmidt}, \citenamefont
  {Shand}, \citenamefont {Kharel}, \citenamefont {Mkhitaryan}, \citenamefont
  {Ning},\ and\ \citenamefont {Ke}}]{lukashev2023jap}%
  \BibitemOpen
  \bibfield  {author} {\bibinfo {author} {\bibfnamefont {P.~V.}\ \bibnamefont
  {Lukashev}}, \bibinfo {author} {\bibfnamefont {A.}~\bibnamefont {Ramker}},
  \bibinfo {author} {\bibfnamefont {B.}~\bibnamefont {Schmidt}}, \bibinfo
  {author} {\bibfnamefont {P.~M.}\ \bibnamefont {Shand}}, \bibinfo {author}
  {\bibfnamefont {P.}~\bibnamefont {Kharel}}, \bibinfo {author} {\bibfnamefont
  {V.}~\bibnamefont {Mkhitaryan}}, \bibinfo {author} {\bibfnamefont
  {Z.}~\bibnamefont {Ning}},\ and\ \bibinfo {author} {\bibfnamefont
  {L.}~\bibnamefont {Ke}},\ }\bibfield  {title} {\bibinfo {title} {{Electronic,
  magnetic, and structural properties of CoVMnSb: Ab initio study}},\
  }\bibfield  {journal} {\bibinfo  {journal} {Journal of Applied Physics}\
  }\textbf {\bibinfo {volume} {134}},\ \href
  {https://doi.org/10.1063/5.0172655} {10.1063/5.0172655} (\bibinfo {year}
  {2023})\BibitemShut {NoStop}%
\bibitem [{\citenamefont {Ke}\ \emph {et~al.}(2013)\citenamefont {Ke},
  \citenamefont {Belashchenko}, \citenamefont {van Schilfgaarde}, \citenamefont
  {Kotani},\ and\ \citenamefont {Antropov}}]{Ke2013}%
  \BibitemOpen
  \bibfield  {author} {\bibinfo {author} {\bibfnamefont {L.}~\bibnamefont
  {Ke}}, \bibinfo {author} {\bibfnamefont {K.~D.}\ \bibnamefont
  {Belashchenko}}, \bibinfo {author} {\bibfnamefont {M.}~\bibnamefont {van
  Schilfgaarde}}, \bibinfo {author} {\bibfnamefont {T.}~\bibnamefont
  {Kotani}},\ and\ \bibinfo {author} {\bibfnamefont {V.~P.}\ \bibnamefont
  {Antropov}},\ }\bibfield  {title} {\bibinfo {title} {Effects of alloying and
  strain on the magnetic properties of {Fe}${}_{16}${N}${}_{2}$},\ }\href
  {https://doi.org/10.1103/physrevb.88.024404} {\bibfield  {journal} {\bibinfo
  {journal} {Physical Review B}\ }\textbf {\bibinfo {volume} {88}},\ \bibinfo
  {pages} {024404} (\bibinfo {year} {2013})}\BibitemShut {NoStop}%
\bibitem [{\citenamefont {Ke}\ \emph {et~al.}(2017)\citenamefont {Ke},
  \citenamefont {Harmon},\ and\ \citenamefont {Kramer}}]{Ke2017}%
  \BibitemOpen
  \bibfield  {author} {\bibinfo {author} {\bibfnamefont {L.}~\bibnamefont
  {Ke}}, \bibinfo {author} {\bibfnamefont {B.~N.}\ \bibnamefont {Harmon}},\
  and\ \bibinfo {author} {\bibfnamefont {M.~J.}\ \bibnamefont {Kramer}},\
  }\bibfield  {title} {\bibinfo {title} {Electronic structure and magnetic
  properties in {T}$_{2}${AlB}$_{2}$ ({T=Fe, Mn, Cr, Co, and Ni}) and their
  alloys},\ }\href {https://doi.org/10.1103/physrevb.95.104427} {\bibfield
  {journal} {\bibinfo  {journal} {Physical Review B}\ }\textbf {\bibinfo
  {volume} {95}},\ \bibinfo {pages} {104427} (\bibinfo {year}
  {2017})}\BibitemShut {NoStop}%
\bibitem [{\citenamefont {Ning}\ \emph {et~al.}(2024)\citenamefont {Ning},
  \citenamefont {Li}, \citenamefont {Tang}, \citenamefont {Banerjee},
  \citenamefont {Fanelli}, \citenamefont {Abernathy}, \citenamefont {Liu},
  \citenamefont {Ueland}, \citenamefont {McQueeney},\ and\ \citenamefont
  {Ke}}]{Ning2024prb}%
  \BibitemOpen
  \bibfield  {author} {\bibinfo {author} {\bibfnamefont {Z.}~\bibnamefont
  {Ning}}, \bibinfo {author} {\bibfnamefont {B.}~\bibnamefont {Li}}, \bibinfo
  {author} {\bibfnamefont {W.}~\bibnamefont {Tang}}, \bibinfo {author}
  {\bibfnamefont {A.}~\bibnamefont {Banerjee}}, \bibinfo {author}
  {\bibfnamefont {V.}~\bibnamefont {Fanelli}}, \bibinfo {author} {\bibfnamefont
  {D.~L.}\ \bibnamefont {Abernathy}}, \bibinfo {author} {\bibfnamefont
  {Y.}~\bibnamefont {Liu}}, \bibinfo {author} {\bibfnamefont {B.~G.}\
  \bibnamefont {Ueland}}, \bibinfo {author} {\bibfnamefont {R.~J.}\
  \bibnamefont {McQueeney}},\ and\ \bibinfo {author} {\bibfnamefont
  {L.}~\bibnamefont {Ke}},\ }\bibfield  {title} {\bibinfo {title} {{Magnetic
  interactions and excitations in ${\mathrm{SrMnSb}}_{2}$}},\ }\href
  {https://doi.org/10.1103/PhysRevB.109.214414} {\bibfield  {journal} {\bibinfo
   {journal} {Phys. Rev. B}\ }\textbf {\bibinfo {volume} {109}},\ \bibinfo
  {pages} {214414} (\bibinfo {year} {2024})}\BibitemShut {NoStop}%
\bibitem [{\citenamefont {Holstein}\ and\ \citenamefont
  {Primakoff}(1940)}]{holstein1940pr}%
  \BibitemOpen
  \bibfield  {author} {\bibinfo {author} {\bibfnamefont {T.}~\bibnamefont
  {Holstein}}\ and\ \bibinfo {author} {\bibfnamefont {H.}~\bibnamefont
  {Primakoff}},\ }\bibfield  {title} {\bibinfo {title} {Field dependence of the
  intrinsic domain magnetization of a ferromagnet},\ }\href
  {https://doi.org/10.1103/PhysRev.58.1098} {\bibfield  {journal} {\bibinfo
  {journal} {Phys. Rev.}\ }\textbf {\bibinfo {volume} {58}},\ \bibinfo {pages}
  {1098} (\bibinfo {year} {1940})}\BibitemShut {NoStop}%
\bibitem [{\citenamefont {Azuah}\ \emph {et~al.}(2009)\citenamefont {Azuah},
  \citenamefont {Kneller}, \citenamefont {Qiu}, \citenamefont
  {Tregenna-Piggott}, \citenamefont {Brown}, \citenamefont {Copley},\ and\
  \citenamefont {Dimeo}}]{Azuah2009}%
  \BibitemOpen
  \bibfield  {author} {\bibinfo {author} {\bibfnamefont {R.~T.}\ \bibnamefont
  {Azuah}}, \bibinfo {author} {\bibfnamefont {L.~R.}\ \bibnamefont {Kneller}},
  \bibinfo {author} {\bibfnamefont {Y.}~\bibnamefont {Qiu}}, \bibinfo {author}
  {\bibfnamefont {P.~L.~W.}\ \bibnamefont {Tregenna-Piggott}}, \bibinfo
  {author} {\bibfnamefont {C.~M.}\ \bibnamefont {Brown}}, \bibinfo {author}
  {\bibfnamefont {J.~R.~D.}\ \bibnamefont {Copley}},\ and\ \bibinfo {author}
  {\bibfnamefont {R.~M.}\ \bibnamefont {Dimeo}},\ }\bibfield  {title} {\bibinfo
  {title} {{DAVE}: A comprehensive software suite for the reduction,
  visualization, and analysis of low energy neutron spectroscopic data},\
  }\href {https://doi.org/10.6028/jres.114.025} {\bibfield  {journal} {\bibinfo
   {journal} {Journal of Research of the National Institute of Standards and
  Technology}\ }\textbf {\bibinfo {volume} {114}},\ \bibinfo {pages} {341}
  (\bibinfo {year} {2009})}\BibitemShut {NoStop}%
\bibitem [{\citenamefont {Filsinger}\ \emph {et~al.}(2017)\citenamefont
  {Filsinger}, \citenamefont {Schnelle}, \citenamefont {Adler}, \citenamefont
  {Fecher}, \citenamefont {Reehuis}, \citenamefont {Hoser}, \citenamefont
  {Hoffmann}, \citenamefont {Werner}, \citenamefont {Greenblatt},\ and\
  \citenamefont {Felser}}]{Filsinger2017}%
  \BibitemOpen
  \bibfield  {author} {\bibinfo {author} {\bibfnamefont {K.~A.}\ \bibnamefont
  {Filsinger}}, \bibinfo {author} {\bibfnamefont {W.}~\bibnamefont {Schnelle}},
  \bibinfo {author} {\bibfnamefont {P.}~\bibnamefont {Adler}}, \bibinfo
  {author} {\bibfnamefont {G.~H.}\ \bibnamefont {Fecher}}, \bibinfo {author}
  {\bibfnamefont {M.}~\bibnamefont {Reehuis}}, \bibinfo {author} {\bibfnamefont
  {A.}~\bibnamefont {Hoser}}, \bibinfo {author} {\bibfnamefont {J.-U.}\
  \bibnamefont {Hoffmann}}, \bibinfo {author} {\bibfnamefont {P.}~\bibnamefont
  {Werner}}, \bibinfo {author} {\bibfnamefont {M.}~\bibnamefont {Greenblatt}},\
  and\ \bibinfo {author} {\bibfnamefont {C.}~\bibnamefont {Felser}},\
  }\bibfield  {title} {\bibinfo {title} {Antiferromagnetic structure and
  electronic properties of {BaCr}$_{2}${As}$_{2}$ and {BaCrFeAs}$_{2}$},\
  }\href {https://doi.org/10.1103/physrevb.95.184414} {\bibfield  {journal}
  {\bibinfo  {journal} {Physical Review B}\ }\textbf {\bibinfo {volume} {95}},\
  \bibinfo {pages} {184414} (\bibinfo {year} {2017})}\BibitemShut {NoStop}%
\bibitem [{\citenamefont {Zeng}\ \emph {et~al.}(2017)\citenamefont {Zeng},
  \citenamefont {Qin}, \citenamefont {Le},\ and\ \citenamefont
  {Hu}}]{Zeng2017prb}%
  \BibitemOpen
  \bibfield  {author} {\bibinfo {author} {\bibfnamefont {J.}~\bibnamefont
  {Zeng}}, \bibinfo {author} {\bibfnamefont {S.}~\bibnamefont {Qin}}, \bibinfo
  {author} {\bibfnamefont {C.}~\bibnamefont {Le}},\ and\ \bibinfo {author}
  {\bibfnamefont {J.}~\bibnamefont {Hu}},\ }\bibfield  {title} {\bibinfo
  {title} {Magnetism and superconductivity in the layered hexagonal transition
  metal pnictides},\ }\href {https://doi.org/10.1103/PhysRevB.96.174506}
  {\bibfield  {journal} {\bibinfo  {journal} {Phys. Rev. B}\ }\textbf {\bibinfo
  {volume} {96}},\ \bibinfo {pages} {174506} (\bibinfo {year}
  {2017})}\BibitemShut {NoStop}%
\bibitem [{\citenamefont {Zhou}\ \emph {et~al.}(2019)\citenamefont {Zhou},
  \citenamefont {Hu}, \citenamefont {Li},\ and\ \citenamefont
  {Wu}}]{Zhou2019jmmm}%
  \BibitemOpen
  \bibfield  {author} {\bibinfo {author} {\bibfnamefont {W.}~\bibnamefont
  {Zhou}}, \bibinfo {author} {\bibfnamefont {P.}~\bibnamefont {Hu}}, \bibinfo
  {author} {\bibfnamefont {S.}~\bibnamefont {Li}},\ and\ \bibinfo {author}
  {\bibfnamefont {S.}~\bibnamefont {Wu}},\ }\bibfield  {title} {\bibinfo
  {title} {First-principles study of the magnetic and electronic properties of
  {ACr$_2$As$_2$ (A=Sr, Ba)}},\ }\href
  {https://doi.org/10.1016/j.jmmm.2018.12.029} {\bibfield  {journal} {\bibinfo
  {journal} {Journal of Magnetism and Magnetic Materials}\ }\textbf {\bibinfo
  {volume} {476}},\ \bibinfo {pages} {254} (\bibinfo {year}
  {2019})}\BibitemShut {NoStop}%
\bibitem [{\citenamefont {Hamri}\ \emph {et~al.}(2021)\citenamefont {Hamri},
  \citenamefont {Djermouni}, \citenamefont {Zaoui},\ and\ \citenamefont
  {Kacimi}}]{Hamri2021jpcs}%
  \BibitemOpen
  \bibfield  {author} {\bibinfo {author} {\bibfnamefont {A.}~\bibnamefont
  {Hamri}}, \bibinfo {author} {\bibfnamefont {M.}~\bibnamefont {Djermouni}},
  \bibinfo {author} {\bibfnamefont {A.}~\bibnamefont {Zaoui}},\ and\ \bibinfo
  {author} {\bibfnamefont {S.}~\bibnamefont {Kacimi}},\ }\bibfield  {title}
  {\bibinfo {title} {Ab-initio study of the electronic structure and magnetic
  properties of {ACr$_2$Pn$_2$ (A =Ba, Sr, Ca, Sc, K; Pn=As, P)}},\ }\href
  {https://doi.org/10.1016/j.jpcs.2020.109850} {\bibfield  {journal} {\bibinfo
  {journal} {Journal of Physics and Chemistry of Solids}\ }\textbf {\bibinfo
  {volume} {150}},\ \bibinfo {pages} {109850} (\bibinfo {year}
  {2021})}\BibitemShut {NoStop}%
\bibitem [{\citenamefont {Mkhitaryan}\ and\ \citenamefont
  {Ke}(2021)}]{mkhitaryan2021prb}%
  \BibitemOpen
  \bibfield  {author} {\bibinfo {author} {\bibfnamefont {V.~V.}\ \bibnamefont
  {Mkhitaryan}}\ and\ \bibinfo {author} {\bibfnamefont {L.}~\bibnamefont
  {Ke}},\ }\bibfield  {title} {\bibinfo {title} {Self-consistently renormalized
  spin-wave theory of layered ferromagnets on the honeycomb lattice},\ }\href
  {https://doi.org/10.1103/PhysRevB.104.064435} {\bibfield  {journal} {\bibinfo
   {journal} {Phys. Rev. B}\ }\textbf {\bibinfo {volume} {104}},\ \bibinfo
  {pages} {064435} (\bibinfo {year} {2021})}\BibitemShut {NoStop}%
\end{thebibliography}%
\bigskip

\end{document}